\documentclass[11pt]{scrartcl}

\usepackage[margin=1in]{geometry}
\usepackage[T1]{fontenc}
\IfFileExists{libertinus.sty}{\usepackage[tt=false]{libertinus}}{\usepackage{lmodern}}
\usepackage{amsmath,amssymb,amsthm,mathtools}
\usepackage{microtype}
\usepackage{enumitem}
\usepackage[dvipsnames]{xcolor}
\usepackage{xurl}
\usepackage[colorlinks,linkcolor=MidnightBlue,citecolor=Plum,urlcolor=blue]{hyperref}
\definecolor{emailgreen}{RGB}{11,102,35}
\newcommand{\emailt}[1]{\href{mailto:#1}{\textcolor{emailgreen}{\texttt{\nolinkurl{#1}}}}}
\usepackage{authblk}
  
\newtheorem{theorem}{Theorem}[section]
\newtheorem{lemma}[theorem]{Lemma}
\newtheorem{proposition}[theorem]{Proposition}

\newtheorem{remark}[theorem]{Remark}
\theoremstyle{definition}

\newcommand{\TV}{d_{\mathrm{TV}}}
\newcommand{\E}{\mathbb{E}}
\newcommand{\Var}{\operatorname{Var}}
\newcommand{\poly}{\operatorname{poly}}

\newcommand{\eps}{\varepsilon}
\newcommand{\indic}{\mathbf{1}}
\newcommand{\bd}{\partial}
\newcommand{\OO}{\widetilde O}

\title{Testing Expansion without Connectivity via Return Probabilities}
\author[]{Nikolas Melissaris\thanks{Supported by the European Research
Council under grant agreement No.~101115790 (OBELiSC).}}
\affil[]{IRIF, CNRS \& Université Paris Cité \\ \emailt{nikolas@irif.fr}}
\date{}

\begin{document}
\maketitle

\begin{abstract}
We study whether every connected component of a bounded-degree graph is an edge expander, given access to its neighbor lists. Dinur and Goldreich give a tester using $\poly(1/\eps)n^\alpha$ queries for every fixed $\alpha>2/3$. We obtain
\[
 \widetilde O_{d,c,\beta}\!\left(\frac{n^{1/2+\beta}}{\eps}
                         +\frac{\sqrt n}{\eps^4}\right)
\]
queries for every fixed $0<\beta\le1$. The tester accepts graphs whose components have edge expansion at least $c$ and rejects graphs $\eps$-far from having edge expansion at least $\kappa\beta c^2/d$ in every component (for an absolute constant $\kappa>0$), each with probability at least $2/3$. No structural promise is imposed on the input.

The tester compares pairwise collision probabilities at two walk lengths and starts further walks at vertices where sampled endpoints coincide. These additional samples control the contribution of vertices with larger return probabilities. We show that the resulting conditions force the longer walk to be nearly uniform on an induced expander with few edges leaving it. For fixed degree $d\ge3$ and completeness threshold $0<c<(d-2\sqrt{d-1})/2$, we show that a sufficiently small rejection expansion threshold of $\gamma d$ requires $n^{1/2+\Omega_d(\gamma)}$ queries when $\eps$ is a sufficiently small multiple of $\gamma$. Thus the dependence of the excess query exponent on the rejection threshold is optimal up to constant factors in this regime. We also prove an $\Omega(\sqrt{n/\log n})$ lower bound for Cayley graphs at infinitely many input sizes.
\end{abstract}

\section{Introduction}\label{sec:intro}

Consider a graph made of two disjoint expanders. There are no edges across the cut separating them, but within either component every set of at most half its vertices has many edges to its complement, relative to the size of the set. Dinur and Goldreich~\cite{DinurGoldreich2025} ask how many queries are needed to test expansion in this sense, with no restriction on the number or sizes of the components.

Fix a constant degree bound $d$. All graphs are simple and undirected. In the bounded-degree model~\cite{GoldreichRon2002}, a query specifies a vertex $v\in[n]$ and an index $i\in[d]$. It returns the $i$th neighbor of $v$, or $\bot$ if $v$ has fewer than $i$ neighbors. The tester knows $n$ and a proximity parameter $0<\eps\le1$. Distance between graphs on $[n]$ is the minimum fraction of entries on which their $nd$-entry neighbor tables differ, over valid neighbor orderings for both graphs.

For a graph $H=(V,E)$ with at least two vertices, its edge expansion is
\[
 h(H)=\min_{\emptyset\ne S\subseteq V,\ |S|\le |V|/2}
 \frac{|E(S,V\setminus S)|}{|S|},
\]
where $E(S,V\setminus S)$ is the set of edges crossing the cut. We set $h(H)=+\infty$ for a one-vertex graph. Let $\Xi_c(n)$ be the set of $n$-vertex graphs of maximum degree at most $d$ in which every component has edge expansion at least $c$, and put $\Xi_c=\bigcup_{n\ge1}\Xi_c(n)$. A graph is $\eps$-far from $\Xi_c(n)$ if reaching any graph in $\Xi_c(n)$ requires changing at least $\eps nd$ neighbor entries.

Random walks expose the difficulty of testing this property. We use a lazy walk that traverses each incident edge with probability $1/(2d)$ at each step and otherwise stays put. One step uses at most one neighbor query: with probability $1/2$ we stay, and otherwise we query a uniform index in $[d]$ and move if it returns a neighbor. The uniform distribution is stationary on each component, including when the graph is not regular. On a connected expander, a sufficiently long walk is nearly uniform on all $n$ vertices. Two independent endpoints then coincide with probability close to $1/n$. We call this event a pairwise collision. Testers for connected expanders use these collisions to check mixing with query complexity close to $\sqrt n$~\cite{GoldreichRon2011,CzumajSohler2010,NachmiasShapira2010,KaleSeshadhri2011}.

If the start lies in an expanding component of size $m$, the collision probability is instead close to $1/m$, and the tester does not know $m$. A sparse cut creates a further difficulty: a walk can spread through one side before it is likely to cross. The set explored by the walk may therefore be smaller than its component. Even knowing the collision probability exactly does not reveal whether the endpoint distribution is close to uniform on some set. For example, the uniform distribution on two points and the distribution $(2/3,1/6,1/6)$ both have collision probability $1/2$.

Dinur and Goldreich first test whether the endpoint distribution is close to uniform on an unspecified set, a task called generalized uniformity testing~\cite{BatuCanonne2017,DiakonikolasKaneStewart2018}. They estimate the size of that set and then test expansion. Their uniformity test compares pairwise collisions with three-way collisions, in which three independent endpoints coincide. For a uniform distribution on $m$ points, constant expected numbers of these events require respectively $\Theta(\sqrt m)$ and $\Theta(m^{2/3})$ samples. The three-way comparison accounts for their query bound of $\poly(1/\eps)n^\alpha$ for every fixed $\alpha>2/3$~\cite{DinurGoldreich2025}.

A collision also identifies the common endpoint. We start further walks from these vertices and compare their return probabilities with the return probability at the original start. Combined with a comparison at two walk lengths, this supplies the information needed to prove expansion without estimating three-way collisions. The algorithm applies to arbitrary bounded-degree graphs.

\subsection{Our results}

Following Dinur and Goldreich~\cite{DinurGoldreich2025}, a pseudo-tester accepts graphs in $\Xi_c(n)$ and rejects graphs $\eps$-far from $\Xi_{c'}(n)$, each with high probability, for a smaller expansion constant $c'$.

\begin{theorem}\label{thm:main}
There is an absolute constant $\kappa>0$ with the following property. Fix an integer $d\ge3$, $0<c\le d$, and $0<\beta\le1$, and put $c_\beta=\kappa\beta c^2/d$. For every $n\ge1$ and $0<\eps\le1$, there is a randomized pseudo-tester using
\[
 \widetilde O_{d,c,\beta}\!\left(\frac{n^{1/2+\beta}}{\eps}
                            +\frac{\sqrt n}{\eps^4}\right)
\]
neighbor queries that accepts every graph in $\Xi_c(n)$ and rejects every graph $\eps$-far from $\Xi_{c_\beta}(n)$, each with probability at least $2/3$.
\end{theorem}

Here $\widetilde O$ suppresses logarithmic factors in $n$ and $1/\eps$. For fixed $d,c,\beta,\eps$, the query bound is $\widetilde O(n^{1/2+\beta})$, with $c_\beta>0$ independent of $n$. This gives exponents arbitrarily close to $1/2$ in the unrestricted problem, as asked by Dinur and Goldreich~\cite[discussion preceding Theorem~1.3]{DinurGoldreich2025}. The parameter $c_\beta$ decreases with $\beta$. Is this loss in expansion necessary? For connected expansion, Chiplunkar et al.~\cite[Theorem~5]{ChiplunkarKapralovKhannaMousavifarPeres2018} give a lower bound with an exponent strictly above $1/2$ for distinguishing expanders from graphs with a sparse balanced cut. In our setting, however, a graph may be close to a union of expanders precisely because it has such a cut. We need to show that the hard instances cannot be repaired cheaply even if the repair disconnects the graph.

\begin{theorem}[Expansion versus query complexity]\label{thm:tradeoff}
Fix an integer $d\ge3$ and $0<c<(d-2\sqrt{d-1})/2$. There are constants $a,b,\gamma_0>0$, depending only on $d,c$, such that, for every fixed $0<\gamma\le\gamma_0$ and $0<\eps\le b\gamma$, every randomized algorithm that distinguishes graphs in $\Xi_c(n)$ from graphs $\eps$-far from $\Xi_{\gamma d}(n)$, with success probability at least $2/3$, uses
\[
 \Omega_{d,c,\gamma}(n^{1/2+a\gamma})
\]
neighbor queries in the worst case, even on $d$-regular inputs, for all sufficiently large $n=2N$ with $dN$ even.
\end{theorem}

For fixed $d,c$ in this range and sufficiently small fixed $\gamma$, take $\beta=\gamma d^2/(\kappa c^2)$ in Theorem~\ref{thm:main}. For fixed $0<\eps\le b\gamma$, the upper and lower bounds give an excess query exponent of $\Theta_{d,c}(\gamma)$ above $1/2$ (up to logarithmic factors in the query complexity). In particular, a $\widetilde O(\sqrt n)$ tester cannot have a positive rejection expansion threshold independent of $\eps$ and work for every $\eps>0$ in this range of $d,c$. The lower bound uses the noisy-parity construction of Chiplunkar et al.~\cite{ChiplunkarKapralovKhannaMousavifarPeres2018}, in the simple-graph formulation of Peng and Shen~\cite{PengShen2026}.

Dinur and Goldreich also prove an $\Omega(\sqrt n)$ lower bound on unrestricted bounded-degree graphs~\cite[Theorem~3.1]{DinurGoldreich2025}. The random lifts in Theorem~\ref{thm:tradeoff} are regular but need not be vertex-transitive. We next ask how much the query complexity can decrease when the input is known to be a Cayley graph. For a finite group $\Gamma$ and a symmetric set $S=S^{-1}$ excluding the identity, the graph $\operatorname{Cay}(\Gamma,S)$ has vertex set $\Gamma$ and edges from $g$ to $gs$ for $s\in S$. The set $S$ need not generate $\Gamma$, so the graph may be disconnected.

\begin{samepage}
\begin{theorem}[Lower bound for Cayley graphs]\label{thm:cayley-lower}
There exist an even constant degree $d\ge6$ and constants $c_+>c_->0$ and $\eps_0>0$ such that, for infinitely many $n$, every randomized algorithm that distinguishes degree-$d$ Cayley graphs (given by neighbor lists with arbitrary vertex labels) in $\Xi_{c_+}(n)$ from degree-$d$ Cayley graphs that are $\eps_0$-far from $\Xi_{c_-}(n)$, with success probability at least $2/3$, uses $\Omega(\sqrt{n/\log n})$ neighbor queries.
\end{theorem}
\end{samepage}

\subsection{Further related work}

The generalized uniformity lower bounds concern algorithms sampling from one unknown distribution~\cite{BatuCanonne2017,DiakonikolasKaneStewart2018}. Our tester also requests samples from walk distributions starting at observed collision locations. Dinur and Goldreich already start further walks at ordinary endpoints~\cite[Section~3.2, Step~2]{DinurGoldreich2025}. A vertex reached with probability $p$ contributes $p^2$ to the collision probability. We use this weighting together with the two return times to obtain the local conditions for the structural proof. Appendix~\ref{app:walkregular} also gives a simpler tester when return probabilities are constant within each component.

Local spectral methods relate Dirichlet eigenvalues on a prescribed region to the conductance of sets contained in that region~\cite{Chung2007LocalCuts}. Kale, Peres, and Seshadhri show that random walks from most vertices of a large expander remain useful after a small arbitrary graph is attached to it, while Chakraborti et al. show that a regular graph in which most vertices mix sufficiently well to the global uniform distribution contains an almost-spanning induced subgraph with expansion bounded in terms of the mixing time~\cite{KalePeresSeshadhri2013,ChakrabortiEtAl2022}. Here the walk may mix on an unknown proper subset of an unknown component. The Cauchy--Schwarz bound for return probabilities also appears in the continuous-time setting of Regev and Shinkar~\cite{RegevShinkar2016}.

Clusterability gives another formulation of graphs with several internally expanding regions. Czumaj, Peng, and Sohler~\cite{CzumajPengSohler2015} require a partition into at most $k$ clusters with large inner conductance and small outer conductance. Chiplunkar et al.~\cite{ChiplunkarKapralovKhannaMousavifarPeres2018} drop the outer-conductance requirement from the acceptance condition and obtain testers with a constant conductance gap. Spectral clustering oracles give local access to cluster structure under suitable clusterability assumptions~\cite{GluchEtAl2021,KapralovKochetkovaWrzosKaminska2026}. In our setting the number of components is unrestricted, and the components may have unrelated sizes.

Peng and Shen~\cite{PengShen2026} ask whether a bounded-degree graph contains an induced expander on a constant fraction of its vertices, allowing arbitrary edges outside the expander and between it and the rest of the graph. They obtain near-square-root query complexity with a logarithmic loss in the conductance guarantee and prove a corresponding lower bound for their structural testing problem. Their tester compares endpoint distributions from different starting vertices; its analysis uses a trace chain that records visits to the expanding region. In our problem even a YES graph may have no induced expander of linear size, and the existence of such an expander says little about the remaining components. Our lower bound uses the binary case of their adaptive indistinguishability lemma, with a separate argument establishing distance from $\Xi_{c'}$.

In degree zero, expansion without connectivity compares the number of edges on which a function changes with its distance from the functions that are constant on each connected component.\footnote{Over $\mathbb F_2$, these componentwise constant functions are the degree-zero cocycles. The distance of $\indic_S$ from this space is $\sum_C\min\{|S\cap C|,|C\setminus S|\}$. With unnormalized counts, componentwise expansion at least $c$ is equivalent to requiring at least $c$ times this distance many edges on which the function changes. Comparing with global constants instead also imposes connectivity.}

\section{Techniques and proof overview}\label{sec:overview}

Let $G=(V,E)$ be the input graph. A walk can spread through a region long before it explores the rest of its component. The tester has to decide whether the region it sees can be retained as an expanding component after few edits. It cannot first identify the component and then run a connected-expansion tester: identifying a large component could already require too many queries.

We follow one starting vertex $s$. The first comparison asks whether doubling the walk length changes its collision probability appreciably. The second compares return probabilities at collision locations with the return probability at $s$, controlling the contribution of vertices with larger return probabilities. These comparisons are the main idea behind the tester. The structural argument shows that they force the longer walk to be nearly uniform on an induced expander with few edges leaving it. When most starts satisfy these conditions, we adapt the overlap argument of Dinur and Goldreich~\cite[Claims~3.4.4--3.4.6]{DinurGoldreich2025} to select disjoint expanding sets covering almost all vertices, giving the graph repair. The repair uses edge deletions alone, so every retained component is an induced expander of the original graph. The sets are used only to prove soundness; the tester estimates return probabilities and never constructs them.

\subsection{Two return times and the vertices where walks collide}\label{sec:overview-test}

Write $P$ for the transition matrix of the lazy walk described in the introduction, so $P^\ell(x,y)$ is the probability of reaching $y$ after $\ell$ steps from $x$. For a walk length $t$, put $p_s=P^{2t}(s,\cdot)$ and
\begin{equation}\label{eq:collision-return}
 A_s=P^{2t}(s,s)=\|P^t(s,\cdot)\|_2^2,
 \qquad B_s=P^{4t}(s,s)=\|p_s\|_2^2.
\end{equation}
These identities follow from symmetry of $P$. Thus $A_s$ and $B_s$ are both return probabilities and collision probabilities: we estimate them by counting pairs of independent endpoints at lengths $t$ and $2t$.

If $s$ lies in an expanding component of size $m$, both distributions are close to uniform on that component for sufficiently large $t$, and $A_s,B_s\approx1/m$. Their ratio is close to one even though the tester does not know $m$. This suggests comparing $A_s$ with $B_s$. To see what such a comparison gives, symmetry and Cauchy--Schwarz imply
\begin{equation}\label{eq:cs}
 p_s(v)^2=\langle P^t(s,\cdot),P^t(v,\cdot)\rangle^2\le A_sA_v.
\end{equation}
If $A_s$ is close to $B_s$, this bounds $p_s(v)$ by about $B_s$ at every vertex with $A_v$ close to $A_s$. Also, $B_s=\sum_vp_s(v)^2$ is the average of $p_s(v)$ when $v$ is drawn from $p_s$. An average close to an upper bound leaves little room for values substantially below that bound. This is the source of near-uniformity in the proof.

Vertices with larger $A_v$ remain a problem: Eq.~\eqref{eq:cs} gives a weaker upper bound there. Sampling a few ordinary endpoints need not detect their importance. A vertex can receive only one percent of a distribution's mass and still account for almost all its collisions, if the rest of the mass is spread over sufficiently many points. We therefore measure how much these vertices contribute to the collision probability. For a sufficiently small accuracy parameter $\eta>0$, define
\begin{equation}\label{eq:large-returns}
 D_s=\{v:A_v>(1+8\eta)A_s\},\qquad
 J_s=\sum_{v\in D_s}p_s(v)^2.
\end{equation}
Call $s$ \emph{good} if
\begin{equation}\label{eq:certificate}
 A_s\le(1+2\eta)B_s,\qquad J_s\le\eta B_s.
\end{equation}
The second condition says that at most an $\eta$ fraction of the collision probability comes from vertices with larger return probabilities. These are conditions used in the soundness proof, with no structural promise on $G$.

The definition suggests how to test the second condition. For any fixed pair of independent $2t$-step walks, the probability that both end at $v$ is $p_s(v)^2$. We take a fresh batch of endpoints and inspect every distinct vertex that appears at least twice. At each such vertex $v$, fresh $t$-step walks estimate $A_v$. If the vertices of $D_s$ contribute enough to the collision probability, the batch should contain a collision at one of them, where the further return estimate detects the larger value.

The batch size is important. After obtaining an estimate of $B_s$ with relative error $O(\eta)$, take
\[
 q=\Theta\!\left(\frac1{\sqrt{\eta B_s}}\right)
\]
fresh endpoints, with a sufficiently large constant. The expected total number of colliding pairs is $\Theta(1/\eta)$. If $J_s>\eta B_s$, the expected number whose common endpoint lies in $D_s$ is bounded below by a large constant. Collision indicators sharing a sample are dependent, so this expectation alone does not establish detection. The second-moment bound in Lemma~\ref{lem:statistics} gives detection with high probability for the chosen batch size. This is why repeated endpoints supply the weighting needed by the proof.

We also reject if the total collision count exceeds a sufficiently large constant times $1/\eta$. Otherwise there are only $O(1/\eta)$ vertices at which to make further return estimates. On an expanding component the cap is exceeded with small probability, while it bounds the query cost on every execution. Section~\ref{sec:collisions} gives the exact comparisons, estimates, and stopping rule. It proves that a start violating Eq.~\eqref{eq:certificate} is rejected with high probability. Completeness uses the stronger fact that, on an expanding component, all the relevant return probabilities are close to $1/m$; we do not require acceptance at every good start.

\subsection{A nearly uniform set with few edges leaving it}\label{sec:overview-set}

Suppose now that $s$ is good. Write $p=p_s$ and $B=B_s$. Outside $D_s$, Eq.~\eqref{eq:cs} gives $p(v)\le(1+O(\eta))B$. Values above $B$ can therefore contribute little excess outside $D_s$, and their contribution inside $D_s$ is bounded by $J_s$. For $X$ drawn from $p$, the random variable $p(X)/B$ has mean one. Its expected excess above one is $O(\eta)$, so its expected deficit below one is the same. Equivalently,
\[
 \sum_v p(v)|p(v)-B|=O(\eta B).
\]
Most of the walk mass must consequently lie at vertices whose endpoint probabilities are close to $B$.

Take the vertices with $p(v)\ge3B/4$. The displayed estimate bounds the probability of landing outside this set by $O(\eta)$, and shows that its size times $B$ is $1\pm O(\eta)$. Removing the vertices of $D_s$ changes its size and probability mass by only $O(\eta)$ relative to these scales. We obtain
\[
 H_s=\{v:p_s(v)\ge3B_s/4\}\setminus D_s,
 \qquad |H_s|B_s=1\pm O(\eta),
\]
and $p_s$ is $O(\eta)$-close in total variation to uniform on $H_s$. Total variation distance is one half of the sum of the absolute differences between the probabilities. Removing $D_s$ also ensures that every remaining vertex satisfies $A_v\le(1+8\eta)A_s$. This return bound will be needed to prove expansion.

Near-uniformity alone does not bound the number of edges leaving $H_s$. The comparison at the two times supplies the missing graph information. For our lazy walk, spectral decomposition gives
\[
 \frac1{2d}\sum_{\{u,v\}\in E}(p_s(u)-p_s(v))^2
 \le A_s-B_s\le2\eta B_s.
\]
Thus the probabilities cannot change sharply across many edges. To use this estimate at the fixed threshold $3B_s/4$, first consider the set before removing $D_s$. An edge leaving it either ends at a vertex with probability below $B_s/2$, or at one with probability between $B_s/2$ and $3B_s/4$. In the first case the probability changes by at least $B_s/4$, so the squared-difference bound limits the number of such edges. In the second case the near-uniformity estimate bounds the number of vertices in that probability range, and each has at most $d$ incident edges. Removing the few vertices in $D_s$ adds few boundary edges. Altogether,
\[
 |E(H_s,V\setminus H_s)|=O(d\eta)|H_s|.
\]
Lemma~\ref{lem:threshold} proves these size, distribution, and boundary estimates together.

\subsection{Internal expansion from return probabilities}\label{sec:overview-expansion}

We have found a region on which the walk is nearly uniform, with few edges leading outside. It could still contain a sparse internal cut. For example, two disconnected expanders have no edges leaving their union, but the union is not an expander. We need to rule out this possibility inside $H_s$.

Once walks have mixed throughout a component of $m$ vertices, each vertex has return probability about $1/m$, so the sum of the return probabilities is about one. Two separate components would each contribute about one. This suggests summing returns inside $H_s$ to detect whether it contains more than one region that retains walks for a long time.

Restrict $P$ to its rows and columns in $H_s$, giving $K_s=P[H_s,H_s]$. An entry of $K_s^\ell$ is the probability of reaching the corresponding endpoint in $\ell$ steps without leaving $H_s$. Its trace sums the return probabilities of these paths. At length $2t$, each diagonal entry is at most the unrestricted return probability $A_v$, and the removal of $D_s$ bounded every such $A_v$. Hence
\[
 \operatorname{tr}(K_s^{2t})
 \le\sum_{v\in H_s}A_v
 \le |H_s|(1+8\eta)A_s
 \le1+O(\eta).
\]

Write $\lambda_1\ge\lambda_2\ge\cdots$ for the eigenvalues of $K_s$. They lie in $[0,1]$, as follows from the laziness of $P$ and restriction to a principal submatrix. The small boundary makes $\lambda_1\ge1-O(\eta)$: a walk started uniformly in $H_s$ leaves it with probability only $O(\eta)$ in one step. Thus $\lambda_1^{2t}$ is at least $1-O(t\eta)$. All other eigenvalues have nonnegative powers, so, when $|H_s|\ge2$,
\[
 \lambda_2^{2t}
 \le\operatorname{tr}(K_s^{2t})-\lambda_1^{2t}
 \le O(t\eta).
\]
Taking logarithms gives $-\log\lambda_2\ge(\log(1/\eta)-\log(Ct))/(2t)$ for an absolute constant $C$ when $\lambda_2>0$. Our parameters make $\log(Ct)$ at most half of $\log(1/\eta)$. Since $t=\Theta(d^2c^{-2}\log(n/\eta))$, a constant gap therefore comes from keeping $\log(1/\eta)/\log(n/\eta)$ bounded below. The choice $\eta\le n^{-\beta/3}$ ensures this for fixed $\beta>0$, even when $\eps$ forces $\eta$ to be smaller. The amount of separation depends on $\beta$, which will determine the parameter $c_\beta$.

The spectral bound first controls cuts of subsets of $H_s$ in the original graph. Some edges counted in such a cut may leave $H_s$ altogether. We therefore still need to pass from those boundary estimates to expansion within an induced subgraph. Since few edges leave $H_s$, Lemma~\ref{lem:trim} shows that removing few vertices suffices. The remaining set $U_s$ induces an expander, has few edges leaving it, and the longer walk is still nearly uniform on it. Lemma~\ref{lem:cores} combines these conclusions, with expansion $\Omega(\beta c^2/d)$ for the parameters chosen below.

\subsection{Disjoint expanding sets and the query bound}\label{sec:overview-combine}

The sets $U_s$ need not be components, and different starts may give overlapping sets. To change the graph into a union of expanders, we have to choose disjoint pieces and show that they cover almost all vertices. Stopping when no more disjoint sets can be added does not yet prove coverage: a set associated with an uncovered start might already meet the selected pieces.

Suppose we have selected several disjoint expanding pieces and consider another set $U$. If its overlap with the selected pieces is small, we can remove that overlap and retain another expander. The difficult case is a set that we can no longer add. We choose the largest available set at each step. Until $U$ first develops too much overlap, every selected piece comes from a set at least as large as $U$.

Now cut each of those pieces into its intersection with $U$ and its complement. The crossing edges leave $U$, so there are few of them in total. Internal expansion forces each cut to have a small side. Once the total overlap is large enough, some selected piece must have almost all its vertices inside $U$. The size comparison then forces almost all of $U$ into that piece as well. This is the reason for selecting the largest available set, and for considering the first stage at which $U$ has too much overlap. Lemma~\ref{lem:combine} makes the argument precise, following the overlap approach of Dinur and Goldreich~\cite[Claims~3.4.4--3.4.6]{DinurGoldreich2025}.

Let $R$ be the vertices left uncovered. The preceding argument bounds the fraction of $U_s$ in $R$ for every good start $s$. Since $p_s$ is nearly uniform on $U_s$, such a start sends only $O_{d,c,\beta}(\eta)$ walk mass into $R$. Let $\rho$ be the fraction of starts that are not good. A walk started uniformly in the graph also ends uniformly, because $P$ is symmetric and stochastic. Averaging over all starts therefore gives
\[
 \frac{|R|}{n}
 =\frac1n\sum_{s\in V}p_s(R)
 \le\rho+O_{d,c,\beta}(\eta).
\]
Delete the edges leaving the selected pieces and all edges incident to $R$. The selected pieces remain induced expanders, and the other vertices become singletons. The number of changed neighbor entries is $O_{d,c,\beta}((\rho+\eta)n)$. Thus, if $\eta$ is a sufficiently small multiple of $\eps$, an $\eps$-far graph must have an $\Omega_{d,c,\beta}(\eps)$ fraction of starts violating the local conditions. Sampling $O_{d,c,\beta}(1/\eps)$ uniform starts finds one with constant probability. Independent repetition at each start reduces the error of the vertex test. Section~\ref{sec:tester} proves these global guarantees.

We can now explain the accuracy and query parameters. Choose a sufficiently small constant $a_0=a_0(d,c,\beta)>0$ and put
\begin{equation}\label{eq:eta}
 \eta=a_0\min\{\eps,n^{-\beta/3}\},
 \qquad t=\Theta\!\left(d^2c^{-2}\log(n/\eta)\right).
\end{equation}
The walk length ensures mixing in every component of expansion at least $c$. The bound $\eta\le a_0\eps$ makes the error in the graph repair small enough. The bound $\eta\le a_0n^{-\beta/3}$ turns the trace estimate into $1-\lambda_2=\Omega(\beta c^2/d^2)$, uniformly even when $\eps$ is smaller than $n^{-\beta/3}$. The spectral and combining arguments give $c_\beta=\Omega(\beta c^2/d)$ as in Theorem~\ref{thm:main}.

Each return estimate uses $\widetilde O(\sqrt n/\eta^2)$ independent endpoints. Since $B_s\ge1/n$, the implementation replaces any estimate below $1/(2n)$ by $1/(2n)$ when choosing the fresh collision batch. This bounds its size by $O(\sqrt{n/\eta})$ even if an estimate fails. The collision cap permits only $O(1/\eta)$ further return estimates, and each walk step uses at most one neighbor query. A trial consequently costs $\widetilde O(\sqrt n\,\eta^{-3}t)$ queries. With $O_{d,c,\beta}(1/\eps)$ starts and logarithmically many repetitions at each start, the total is
\[
 \widetilde O_{d,c,\beta}\!\left(\frac{\sqrt n}{\eps\eta^3}\right)
 =\widetilde O_{d,c,\beta}\!\left(\frac{n^{1/2+\beta}}\eps+
                                    \frac{\sqrt n}{\eps^4}\right).
\]
The exponent above $1/2$ pays for the accuracy needed to prove internal expansion. The same choice of $\eta$ controls this extra query exponent and the expansion threshold $c_\beta$.

\subsection{Why the rejection threshold affects the query exponent}\label{sec:overview-tradeoff}

Take two copies of a random regular graph. For each base edge, either keep its two copies within their respective layers or replace them by two edges between the layers. This gives a $2$-lift of the base graph. If the choices are independent uniform bits, the lift is an expander with high probability. For the NO graph, we use independent noise bits of small mean $p$ to decide which edges cross between layers, and then independently permute the two vertices above each base vertex. The noisy-parity lower bound~\cite{ChiplunkarKapralovKhannaMousavifarPeres2018,PengShen2026} says that distinguishing these two distributions requires $n^{1/2+\Omega_d(p)}$ queries.

The NO graph has a balanced cut of expansion $O(pd)$, given by the layers before their vertices were permuted. We still need to show that this graph is far from $\Xi_{c'}$, rather than merely failing to expand by $c'$. To do so, we prove that its own expansion is $\Omega_d(p)$. A cut choosing one vertex above every base vertex corresponds to a choice of which fibers to swap. Swapping a set of fibers toggles precisely the edges in its base cut. To eliminate a substantial fraction of the noisy edges, the connected pieces of the swapped set would have to account for a large total excess of noisy over non-noisy boundary edges. Base expansion and independence of the noise make the total improvement from all such pieces small. For an arbitrary cut, base expansion controls the number of fibers where both or neither vertex is chosen. If at least half the fibers are empty, base expansion directly bounds the cut; otherwise we reduce to a cut choosing one vertex from every fiber. This gives the desired expansion bound for the NO graph.

With $p=\Theta(c'/d)$ and a sufficiently small proportionality constant, the balanced cut has expansion at most $c'/2$, while the graph still has expansion $\Omega_d(c')$. The farness argument in Lemma~\ref{lem:farness} then shows that any repair requires $\Omega_d(c'n)$ edge modifications. Section~\ref{sec:tradeoff} makes these arguments precise.

\subsection{Cayley graphs that require many queries}\label{sec:overview-lower}

For our second lower bound, the graph has to remain hard even when the tester knows it is a Cayley graph. Start with a bipartite Lubotzky--Phillips--Sarnak (LPS) Ramanujan Cayley graph. For these $d$-regular graphs, the Ramanujan bound says that every adjacency eigenvalue other than $\pm d$ has absolute value at most $2\sqrt{d-1}$. The YES graph is a fixed number $L$ of disjoint copies of this base graph. Label the copies cyclically. In the NO graph, traversing any of the chosen representatives of the inverse pairs increases the copy index by one, and traversing its inverse decreases it by one. The resulting graph is connected, and both graphs project to the same LPS base after ignoring the copy index.

A consecutive block of copies gives a balanced cut in the NO graph with expansion below the target. This alone would not prove farness: deleting a sparse cut could produce expanding components. We also show that the NO graph retains a fixed positive expansion lower bound. Any repair using few edits must then leave almost all vertices in one component. That component still meets both sides of the balanced cut substantially, so meeting the target expansion requires many added crossing edges. This proves constant distance from the target property even though repairs may disconnect the graph.

It remains to hide the difference between the two graphs from an adaptive query algorithm. Randomize the vertex names and neighbor-list orderings. An algorithm can branch from a previously seen vertex, start a new exploration at an arbitrary label, and compare the answers. We compare its whole transcript with an auxiliary oracle that answers as a forest of $d$-regular trees. Until two distinct forest vertices are mapped to the same graph vertex, the answers can agree with this oracle in either hard graph.

The forest is generated before assigning hidden group generators to its edges. Conditional on this forest and its transcript, the path between any two vertices in the same tree has a uniformly random sequence of generators with no immediate inverse pair. Its projection is therefore a non-backtracking walk in the LPS graph, meaning a walk that never immediately retraces its last edge. For the vertices to coincide in either hard graph, this projected walk must return to its start.

The absence of short cycles rules out such returns for short paths. An adaptive tester can also create longer paths, so we need a return bound at every length. Arithmetic counting controls the intermediate lengths, and the Ramanujan eigenvalue bound controls the longer ones. Together they give a return probability $O(\log n/n)$ uniformly in the path length. Vertices in different trees coincide with probability $O(1/n)$. A $Q$-query exploration exposes only $O(Q)$ vertices, so a union bound over all pairs gives transcript distance $O(Q^2\log n/n)$. For $Q$ a sufficiently small constant times $\sqrt{n/\log n}$, the two graphs cannot be distinguished with probability $2/3$.

Section~\ref{sec:lower} gives the construction, applies the common farness criterion, and proves the uniform return estimate before applying them to the adaptive transcript. In particular, it includes the arithmetic estimate responsible for the logarithmic loss.

\section{Preliminaries}\label{sec:prelim}

Throughout the upper-bound proof, $d,c,\beta$ are as in Theorem~\ref{thm:main}, and $\eta=a_0\min\{\eps,n^{-\beta/3}\}$ as in Eq.~\eqref{eq:eta}. We use the walk length
\begin{equation}\label{eq:walk-time}
 t=\left\lceil16d^2c^{-2}\log(n/\eta)\right\rceil.
\end{equation}
We choose $a_0=a_0(d,c,\beta)>0$ sufficiently small to satisfy the structural and repair bounds below. We write $C(s)$ for the component containing $s$, $\partial W=E(W,V\setminus W)$ for the edge boundary of a vertex set $W$, and $u_W$ for its uniform distribution when $W$ is nonempty. For distributions $p,q$ on a finite set, $\TV(p,q)=\frac12\sum_v|p(v)-q(v)|$.

\subsection{Return probabilities}

The transition matrix of the walk described in the introduction is
\[
 P(x,y)=\begin{cases}
 1/(2d),&\{x,y\}\in E,\\
 1-\deg(x)/(2d),&x=y,\\
 0,&\text{otherwise.}
 \end{cases}
\]
Equivalently, $P=I-L_G/(2d)$, where $L_G$ is the graph Laplacian. The eigenvalues of $L_G$ lie in $[0,2d]$, so all eigenvalues of $P$ lie in $[0,1]$. The matrix is symmetric and stochastic, and its restriction to each component is doubly stochastic: every row and column sums to one. For the quantities in Eq.~\eqref{eq:collision-return}, Cauchy--Schwarz and spectral decomposition give
\begin{equation}\label{eq:return-range}
 1/n\le B_s\le A_s\le1.
\end{equation}

\begin{lemma}[Mixing]\label{lem:mixing}
Let $C$ be a component of size $m$ with $h(C)\ge c$. For every $x\in C$ and integer $\ell\ge0$,
\[
 \|P^\ell(x,\cdot)-u_C\|_2^2\le \exp(-c^2\ell/(4d^2)).
\]
In particular, the choice of $t$ in Eq.~\eqref{eq:walk-time}, with $a_0$ sufficiently small, ensures
\begin{equation}\label{eq:mixed}
 1/m\le A_v,B_v\le(1+\eta/10)/m\qquad(v\in C).
\end{equation}
\end{lemma}

\begin{proof}
If $m=1$, the assertions are immediate. Otherwise, the probability that a walk started uniformly in a nonempty $S\subseteq C$, $|S|\le m/2$, leaves $S$ in one step is $|\partial S|/(2d|S|)\ge c/(2d)$. Cheeger's inequality gives a spectral gap at least $c^2/(8d^2)$; see, e.g.,~\cite{HooryLinialWigderson2006}. Expanding the point mass at $x$ in an orthonormal eigenbasis proves the first bound. Orthogonality to the stationary vector gives
\[
 A_v=1/m+\|P^t(v,\cdot)-u_C\|_2^2,
 \qquad B_v=1/m+\|P^{2t}(v,\cdot)-u_C\|_2^2.
\]
The first error is at most $(\eta/n)^4\le\eta/(10n)\le\eta/(10m)$, and the second is no larger.
\end{proof}

\section{The tester}\label{sec:collisions}

We now specify the vertex test described in Section~\ref{sec:overview-test}. We first record the collision estimates used to choose its samples and detect a collision in $D_s$. They hold for arbitrary distributions.

\begin{lemma}[Collision estimates]\label{lem:statistics}
Let $r$ be a distribution on at most $n$ points, $B=\sum_vr(v)^2$, and $q\ge2$. From $q$ independent samples define $N_v$ as the multiplicity of $v$ and $S=\sum_v\binom{N_v}2$.
\begin{enumerate}[label=(\roman*)]
\item Given $0<u<1$ and $0<\zeta\le1/3$, one can estimate $B$ to relative error $u$, with failure probability at most $\zeta$, using
\[
 O(\sqrt n\,u^{-2}\log(1/\zeta))
\]
samples.
\item For any fixed set $D$, put $J=\sum_{v\in D}r(v)^2$. For every $K>0$, if $q\sqrt J\ge K$, the probability that no repeated sample lies in $D$ is $O(K^{-1}+K^{-2})$.
\end{enumerate}
\end{lemma}

\begin{proof}
For $S_D=\sum_{v\in D}\binom{N_v}2$, disjoint sample pairs give independent indicators. Each triple of sample indices gives three pairs of overlapping indicators, whose joint event is that all three samples coincide in $D$. Expanding the variance therefore gives
\[
 \E S_D=\binom q2 J,\qquad
 \Var S_D=\binom q2(J-J^2)+6\binom q3\left(\sum_{v\in D}r(v)^3-J^2\right).
\]
Thus, using $\sum_{v\in D}r(v)^3\le J^{3/2}$,
\[
 \Var S_D\le \tfrac12q^2J+q^3J^{3/2}.
\]
For $D$ equal to the whole domain, normalizing by $\binom q2$ gives an unbiased estimate $\widehat B$ with
\[
 \Var\widehat B=O(B/q^2+B^{3/2}/q).
\]
Since $B\ge1/n$, taking $q=\lceil K\sqrt n/u^2\rceil$ makes the probability of relative error greater than $u$ at most $O(K^{-2}+K^{-1})$ by Chebyshev's inequality. Choose a sufficiently large absolute $K$ so that this probability is at most $1/10$. Taking a median of an odd number of independent batches, using $O(\log(1/\zeta))$ batches, proves (i). For (ii), Chebyshev and $\binom q2\ge q^2/4$ give
\[
 \Pr[S_D=0]\le\frac{\Var S_D}{(\E S_D)^2}
 =O\!\left(\frac1{q^2J}+\frac1{q\sqrt J}\right).
\]
\end{proof}

Fix sufficiently large absolute constants $K$ and $C_{\mathrm{cap}}$. We now give one trial of the test at a starting vertex $s$. All estimates and walks use fresh independent randomness conditional on the previous transcript. Their starting vertices and batch sizes may depend on that transcript.
\begin{enumerate}
\item Estimate $A_s$ and $B_s$ to relative accuracy $u=\eta/100$, each with failure probability at most $1/100$. The estimates use endpoints of $t$-step and $2t$-step walks, respectively. Reject if $\widehat A_s>(1+\eta)\widehat B_s$.
\item Set $\widetilde B_s=\max\{\widehat B_s,1/(2n)\}$ and take a fresh batch of
\[
 q=\left\lceil K/\sqrt{\eta\widetilde B_s}\right\rceil
\]
independent $2t$-step endpoints from $s$. Compute $S=\sum_v\binom{N_v}2$ from their multiplicities. If $S>M=\lceil C_{\mathrm{cap}}K^2/\eta\rceil$, reject.
\item For every distinct vertex $y$ with $N_y\ge2$, estimate $A_y$ to relative accuracy $u$, using fresh $t$-step walks from $y$, with failure probability at most $1/(100M)$. Reject if any such estimate satisfies
\[
 \widehat A_y>(1+4\eta)\widehat A_s.
\]
Accept otherwise.
\end{enumerate}
There are at most $M$ vertices in the last step whenever it is reached, since each distinct repeated endpoint contributes at least one collision. Multiplicities can be computed by sorting or hashing; sample pairs are not enumerated.

The full tester uses the following procedure, for a sufficiently large constant $C_{\mathrm{start}}=C_{\mathrm{start}}(d,c,\beta)$.
\begin{enumerate}
\item Sample $N_{\mathrm{start}}=\lceil C_{\mathrm{start}}/\eps\rceil$ vertices independently and uniformly from $[n]$.
\item At each sampled vertex, run an odd number $\Theta(\log(100N_{\mathrm{start}}))$ of independent trials of the one-vertex test and take their majority decision. The absolute constant in the number of trials is sufficiently large.
\item Reject if any majority decision rejects; accept otherwise.
\end{enumerate}
Section~\ref{sec:tester} justifies the choice of $C_{\mathrm{start}}$ and proves the guarantees of the full tester.

\begin{proposition}[One-vertex test]\label{prop:vertextest}
A trial uses $\OO(\sqrt n\,\eta^{-3}t)$ neighbor queries on every execution. It rejects a vertex violating the local conditions~\eqref{eq:certificate} with probability at least $9/10$. If its component satisfies Eq.~\eqref{eq:mixed}, it accepts with probability at least $9/10$.
\end{proposition}

\begin{proof}
The first two estimates are simultaneously accurate except with probability $2/100$. On this event the lower cutoff does not alter $\widehat B_s$, and $B_s/2\le\widehat B_s\le2B_s$. If $A_s>(1+2\eta)B_s$, relative accuracy $u=\eta/100$ forces rejection in step 1.

Suppose instead that $J_s>\eta B_s$. Then $q\sqrt{J_s}\ge K/\sqrt2$. By Lemma~\ref{lem:statistics}(ii), choosing $K$ large enough ensures the fresh batch contains a repeated vertex in $D_s$ except with probability $1/100$. If the cap is exceeded, rejection already occurs. Otherwise every collision location is inspected, and all its return estimates are accurate except with probability $1/100$, by conditioning on the batch and applying a union bound over at most $M$ locations. For $y\in D_s$, accurate estimates imply
\[
 \widehat A_y/\widehat A_s
 \ge\frac{1-u}{1+u}(1+8\eta)>1+4\eta,
\]
so step 3 rejects. This proves the rejection guarantee.

Under \eqref{eq:mixed}, accurate initial estimates cause step 1 to accept, and accurate estimates at every inspected vertex cause step 3 to accept. It remains to bound the probability that $S>M$. Conditional on accurate initial estimates, $q\le2K/\sqrt{\eta B_s}$, so
\[
 \E S=\binom q2 B_s\le2K^2/\eta,
 \qquad\Pr[S>M]\le2/C_{\mathrm{cap}}\le1/100.
\]
Together with the estimate failure bounds, this proves acceptance with probability greater than $9/10$.

For worst-case query complexity, the lower cutoff ensures $q=O(\sqrt{n/\eta})$. The first two estimates cost $\OO(\sqrt n\eta^{-2}t)$ queries. At most $M=O(\eta^{-1})$ further estimates are made. Each uses
\[
 O(\sqrt n\eta^{-2}t\log(100M))
\]
queries. Thus the query bound holds in the worst case.
\end{proof}

\section{Expanding sets from return probabilities}\label{sec:structural}

For a good start, we first find a set on which the longer walk is nearly uniform and that has few edges leaving it. The return bound on its vertices also controls the trace of the restricted walk. We then use this trace bound to prove internal expansion after removing few vertices.

\subsection{Near-uniformity, boundary size, and return probabilities}

\begin{lemma}[A nearly uniform set with small boundary]\label{lem:threshold}
There is an absolute constant $C$ such that, for all sufficiently small $\eta>0$, every good start $s$ has a nonempty set
\[
 H_s=\{v:p_s(v)\ge3B_s/4\}\setminus D_s
\]
satisfying
\[
 |H_s|B_s=1\pm C\eta,\qquad
 \TV(p_s,u_{H_s})\le C\eta,\qquad
 |E(H_s,V\setminus H_s)|\le Cd\eta |H_s|,
\]
and, with $K_s=P[H_s,H_s]$,
\begin{equation}\label{eq:trace}
 \operatorname{tr}(K_s^{2t})\le1+C\eta.
\end{equation}
\end{lemma}

\begin{proof}
Write $p=p_s$, $A=A_s$, and $B=B_s$. For $v\notin D_s$, Eq.~\eqref{eq:cs} and Eq.~\eqref{eq:certificate} give
\[
 p(v)\le\sqrt{1+8\eta}\,A
 \le\sqrt{1+8\eta}(1+2\eta)B=(1+O(\eta))B.
\]
For $X\sim p$, put $Z=p(X)/B$, and write $x_+=\max\{x,0\}$. Then $\E Z=1$, and
\[
 \E[(Z-1)_+]
 \le O(\eta)+\frac1B\sum_{v\in D_s}p(v)^2=O(\eta).
\]
The expected deficit equals the expected excess. Hence
\begin{equation}\label{eq:weighted}
 \sum_vp(v)|p(v)-B|=B\E|Z-1|=O(\eta B).
\end{equation}
Let $H_0=\{v:p(v)\ge3B/4\}$. The deficit bound implies $p(H_0^c)=O(\eta)$. Also,
\[
 \sum_{v\in H_0}|p(v)-B|
 \le\frac4{3B}\sum_vp(v)|p(v)-B|=O(\eta).
\]
Consequently $B|H_0|=1\pm O(\eta)$ and $\TV(p,u_{H_0})=O(\eta)$. On $H_0$, the inequalities $p(v)^2\ge9B^2/16$ and $p(v)\le4p(v)^2/(3B)$, together with $J_s\le\eta B$, give
\[
 |H_0\cap D_s|=O(\eta/B),\qquad p(H_0\cap D_s)=O(\eta).
\]
Removing these vertices leaves $H_s$, with $B|H_s|=1\pm O(\eta)$ and $\TV(p,u_{H_s})=O(\eta)$. In particular, $H_s$ is nonempty for sufficiently small $\eta$.

For the boundary, let $(v_j)_j$ be an orthonormal eigenbasis of $P$, with $Pv_j=\lambda_jv_j$, and let $e_s$ be the unit vector at $s$. Then
\[
 \langle p,(I-P)p\rangle
 =\sum_j\lambda_j^{4t}(1-\lambda_j)|\langle e_s,v_j\rangle|^2
 \le\sum_j\lambda_j^{2t}(1-\lambda_j^{2t})|\langle e_s,v_j\rangle|^2
 =A-B\le2\eta B.
\]
For our walk, $\sum_{\{u,v\}\in E}(p(u)-p(v))^2=2d\langle p,(I-P)p\rangle=O(d\eta B)$. Every edge from $H_0$ to $\{v:p(v)<B/2\}$ contributes at least $B^2/16$, so there are $O(d\eta/B)$ such edges. The intermediate set $\{B/2\le p<3B/4\}$ has mass $O(\eta)$ and size $O(\eta/B)$, so at most $O(d\eta/B)$ edges end there. Removing $H_0\cap D_s$ adds at most $d|H_0\cap D_s|=O(d\eta/B)$ boundary edges. Since $B|H_s|=1\pm O(\eta)$, this proves $|\bd H_s|=O(d\eta)|H_s|$.

Finally, every $v\in H_s$ has $A_v\le(1+8\eta)A_s$. A walk counted by $K_s^{2t}$ remains in $H_s$, so $(K_s^{2t})(v,v)\le P^{2t}(v,v)=A_v$. Therefore
\[
 \operatorname{tr}(K_s^{2t})
 \le |H_s|(1+8\eta)A_s
 \le (1+O(\eta))(1+8\eta)(1+2\eta)
 =1+O(\eta).
\]
This proves the trace bound without bounding return probabilities outside $H_s$. All implicit constants in this proof are absolute.
\end{proof}

\subsection{From a spectral bound to an induced expander}

\begin{lemma}[Expansion of subsets in $G$]\label{lem:ambient-expansion}
Let $H$ have size $m\ge 2$, put $K=P[H,H]$, and set
\[
    b=\frac{|E(H,V\setminus H)|}{2dm}.
\]
Suppose that $g>0$, $\lambda_2(K)\le 1-g$, and $b\le g/100$.
Then every nonempty $S\subseteq H$ with $|S|\le m/2$ satisfies
\[
 |E(S,V\setminus S)|\ge \frac{dg}{2}|S|.
\]
\end{lemma}

\begin{proof}
Let $v_1$ be a unit top eigenvector of $K$, chosen nonnegative, and let $u=m^{-1/2}\mathbf1_H$. Since $u^\top Ku=1-b$, while all other eigenvalues are at most $1-g$, we have $1-b\le1-g(1-|\langle u,v_1\rangle|^2)$, and hence $1-|\langle u,v_1\rangle|^2\le b/g$. For $S\subseteq H$ with $r=|S|/m\le1/2$, let $f=|S|^{-1/2}\mathbf1_S$. Write $v_1=\alpha u+w$ with $w\perp u$. The preceding bound gives $\|w\|\le\sqrt{b/g}$, while $\langle f,u\rangle=\sqrt r$. Hence Cauchy--Schwarz gives
\[
 |\langle f,v_1\rangle|
 \le \sqrt r+\sqrt{b/g}.
\] Under $b/g\le1/100$, the square of the right-hand side is at most $3/4$. Spectral decomposition then gives
\[
 \mathbf1_S^\top(I-K)\mathbf1_S
 \ge \frac g4|S|.
\]
For the lazy walk, $\mathbf1_S^\top(I-K)\mathbf1_S=|E(S,V\setminus S)|/(2d)$, which proves the claim.
\end{proof}

The preceding lemma counts all edges leaving $S$ in the original graph. Some of these edges may leave $H$ altogether, so the bound does not yet prove expansion inside $G[H]$. A small set $S$ could have few edges to the rest of $H$ even when its boundary in $G$ is large. The next lemma uses the small boundary of $H$ to show that removing few vertices leaves an induced expander.

\begin{lemma}[Trimming to an induced expander]\label{lem:trim}
Let $G$ have maximum degree at most $d$, let $H\subseteq V(G)$
have size $m\ge 1$, and fix $0<h\le d$. Suppose that every
$S\subseteq H$ with $|S|\le m/2$ satisfies
\[
 |E(S,V\setminus S)|\ge h|S|,
\]
and that
\[
 |E(H,V\setminus H)|\le\tau m,
 \qquad
 \tau\le\frac{h^2}{100d}.
\]
Then there is $U\subseteq H$ such that
\[
 |H\setminus U|\le \frac{2\tau}{h}m,
 \qquad
 h(G[U])\ge h/4,
\]
and
\[
 |E(U,V\setminus U)|\le C\frac{d\tau}{h}|U|
\]
for an absolute constant $C$.
\end{lemma}

\begin{proof}
Put $\alpha=h/4$. Among all $R\subseteq H$ with $|R|\le m/2$, choose one maximizing $F(R)=\alpha|R|-|E(R,H\setminus R)|$. Since $F(\emptyset)=0$, the maximizer satisfies $F(R)\ge0$. Applying the assumed boundary lower bound to $R$ gives $h|R|\le |E(R,H\setminus R)|+|E(R,V\setminus H)|$, so $0\le F(R)\le\tau m-(h-\alpha)|R|$. Hence $|R|\le(4\tau/3h)m\le(2\tau/h)m$. Set $U=H\setminus R$.

Suppose for contradiction that some $S\subseteq U$ with $|S|\le|U|/2$ satisfies $|E(S,U\setminus S)|<\alpha|S|$. If $|R|+|S|\le m/2$, then
\begin{align*}
 F(R\cup S)-F(R)
 &=\alpha|S|-|E(S,U\setminus S)|+|E(S,R)|\\
 &>0,
\end{align*}
contradicting maximality. Thus $|R|+|S|>m/2$. The bound $\tau\le h^2/(100d)$ and $h\le d$ imply $|R|\le m/50$, so $|S|>m/3$. Applying the same boundary lower bound to $S$ now yields
\[
 (h-\alpha)|S|
 <|E(S,R)|+|E(S,V\setminus H)|
 \le d|R|+\tau m,
\]
whereas the right-hand side is at most
\[
 \left(\frac{2d\tau}{h}+\tau\right)m
 \le \frac{3h}{100}m,
\]
and the left-hand side is greater than $hm/4$. This contradiction proves $h(G[U])\ge h/4$.

Finally,
\[
 |E(U,V\setminus U)|
 \le |E(H,V\setminus H)|+d|R|
 \le \left(\tau+\frac{2d\tau}{h}\right)m.
\]
Since $|U|=(1-O(\tau/h))m$, the final bound follows after adjusting the constant.
\end{proof}

We can now obtain the required expanding set. The bound $\eta\le a_0n^{-\beta/3}$ in Eq.~\eqref{eq:eta} turns the trace estimate into a constant spectral gap at the chosen walk length $t$.

\begin{lemma}[An expanding set from a good start]\label{lem:cores}
For a sufficiently small absolute constant $\kappa>0$, put $g=32\kappa\beta c^2/d^2$ and $h_*=dg/8$. If $a_0=a_0(d,c,\beta)>0$ in Eq.~\eqref{eq:eta} is sufficiently small, every good start $s$ has a nonempty set $U_s\subseteq C(s)$ satisfying
\[
 h(G[U_s])\ge h_*,\qquad
 |\partial U_s|\le C_1\eta|U_s|,\qquad
 \TV(p_s,u_{U_s})\le C_1\eta,
\]
for a constant $C_1>0$ depending only on $d,c,\beta$.
\end{lemma}

\begin{proof}
Apply Lemma~\ref{lem:threshold}. If $|H_s|=1$, take $U_s=H_s$ and use the singleton convention. Otherwise put $K=P[H_s,H_s]$ and $b=|\bd H_s|/(2d|H_s|)=O(\eta)$. Its eigenvalues satisfy $1\ge\lambda_1\ge\lambda_2\ge\cdots\ge0$. The Rayleigh quotient of the normalized all-ones vector is $1-b$, so $\lambda_1\ge1-b$. For sufficiently small $a_0$, we have $b<1$. Bernoulli's inequality and the trace bound yield, for an absolute constant $C_2$,
\[
 \lambda_2^{2t}\le1+O(\eta)-(1-b)^{2t}\le C_2t\eta.
\]
To extract a spectral gap uniformly in $n$ and $\eps$, put $L=\log(1/\eta)$ and $x=\log n$. Since $a_0\le1$ and $\eta\le a_0n^{-\beta/3}$, we have $L\ge(\beta/3)x$. For $L\ge1$, Eq.~\eqref{eq:walk-time} also gives $t\le D(x+L)$ for a constant $D=O(d^2/c^2)$, with an absolute implicit constant. Thus
\[
 \log(C_2t)\le\log\bigl(C_2D(1+3/\beta)\bigr)+\log L.
\]
Choose $a_0$ small enough that $\log(C_2D(1+3/\beta))\le L/4$ and $\log L\le L/4$ for every $L\ge\log(1/a_0)$. These conditions depend only on $d,c,\beta$. They imply $\log(1/(C_2t\eta))\ge L/2$. If $\lambda_2>0$, taking logarithms gives
\[
 -\log\lambda_2\ge\frac{L}{4t}
 =\Omega(\beta c^2/d^2).
\]
Here $L/(x+L)\ge\beta/(3+\beta)\ge\beta/4$. Choose the absolute constant $\kappa$ small enough that the last bound is at least $2g$ and $g\le1/2$. Then $1-e^{-2g}\ge g$, so $1-\lambda_2\ge g$. If $\lambda_2=0$, the same conclusion is immediate.

Decrease $a_0$ further so that $b\le g/100$ and
\[
 |\bd H_s|/|H_s|\le h_0^2/(100d),\qquad h_0=dg/2.
\]
This is possible because $b=O(\eta)$ and $|\bd H_s|/|H_s|=O(d\eta)$. Lemma~\ref{lem:ambient-expansion} gives $|\partial S|\ge h_0|S|$ for every $S\subseteq H_s$ with $|S|\le|H_s|/2$. Lemma~\ref{lem:trim} then gives an induced $h_0/4$-expander $U_s\subseteq H_s$ with
\[
 |H_s\setminus U_s|=O(\eta/g)|H_s|,\qquad
 |\bd U_s|=O(d\eta/g)|U_s|.
\]
The total-variation bound follows from
\[
 \TV(p_s,u_{U_s})\le\TV(p_s,u_{H_s})+
                 |H_s\setminus U_s|/|H_s|
 =O(\eta+\eta/g).
\]
These bounds give a common constant $C_1=C_1(d,c,\beta)$, also covering the singleton case. All restrictions on $a_0$ depend only on $d,c,\beta$ and hold for every $n\ge1$ and $0<\eps\le1$. The constants are unchanged by any later decrease of $a_0$.
\end{proof}

\section{Combining the sets and proving soundness}\label{sec:tester}

The sets from good starts must cover almost all vertices after small overlaps are removed. The next lemma formalizes the selection in Section~\ref{sec:overview-combine}. It bounds the repair cost linearly in the fraction of starts where the local conditions fail, without requiring sets at those starts. The proof follows the overlap approach of Dinur and Goldreich~\cite[Claims~3.4.4--3.4.6]{DinurGoldreich2025}, using the largest available set and the first stage at which another set has too much overlap.

\begin{lemma}[Combining expanding sets]\label{lem:combine}
Fix $d$ and $0<h\le d$. There are $\tau_0,C>0$, depending only on $d,h$, such that the following holds. Let $C_0$ be a component, let $Q$ be a doubly stochastic matrix on it, and let $0\le\rho\le1$. Suppose a fraction at least $1-\rho$ of its vertices $s$ have nonempty sets $U_s\subseteq C_0$ satisfying
\[
 h(G[U_s])\ge h,\quad |\bd U_s|\le\tau|U_s|,
 \quad\TV(Q(s,\cdot),u_{U_s})\le\tau,\quad 0<\tau\le\tau_0.
\]
Then the graph on $C_0$ can be changed into a graph in $\Xi_{h/4}$ by edge deletions using at most $C(\rho+\tau)|C_0|$ changed neighbor entries. 
\end{lemma}

\begin{proof}
Put $a=8/h$ and choose $\tau_0>0$ sufficiently small, depending only on $d,h$. Construct disjoint sets $V_i$ greedily. With $W$ their current union, call $U_s$ admissible if $|U_s\cap W|\le a\tau|U_s|$. Choose an admissible set of maximum size, call it $U_i$, and apply Lemma~\ref{lem:trim} to $H_i=U_i\setminus W$. Indeed, every subset of $H_i$ of size at most $|H_i|/2$ has edge boundary in $G$ at least $h$ times its size, $|H_i|\ge|U_i|/2$, and $|\bd H_i|\le(1+da)\tau|U_i|$. For sufficiently small $\tau_0$, the trimming hypothesis holds and yields a nonempty $V_i\subseteq H_i$ with
\[
 h(G[V_i])\ge h/4,\quad |U_i\setminus V_i|\le b\tau|U_i|,
 \quad |\bd V_i|\le K_0\tau|V_i|.
\]
Here $b,K_0>0$ are constants depending only on $d,h$. The size estimate uses the overlap loss and the trimming loss $2|\bd H_i|/h$; the boundary estimate follows from $|\bd V_i|\le|\bd H_i|+d|H_i\setminus V_i|$ and $|V_i|\ge|U_i|/2$. Choose $\tau_0$ small enough that $a\tau_0\le1/2$, $b\tau_0\le1/2$, and the trimming hypothesis holds. Continue until no admissible set remains, and let $R$ be the uncovered vertices. Each iteration adds a nonempty disjoint set, so at most $|C_0|$ iterations are possible.

For any one of the given sets $U$, put $W_j=\bigcup_{i\le j}V_i$ and $W_0=\emptyset$. Let $j$ be the first index with $|U\cap W_j|>a\tau|U|$. Thus $U$ was admissible immediately before every selection $i\le j$, so maximality gives $|U_i|\ge|U|$. As the $V_i$ are disjoint,
\[
 \frac h4\sum_{i\le j}\min\{|V_i\cap U|,|V_i\setminus U|\}
 \le|\bd U|\le\tau|U|.
\]
But $|U\cap\bigcup_{i\le j}V_i|>8\tau|U|/h$. Therefore some $i\le j$ has $|V_i\cap U|>|V_i\setminus U|$. Its complement in $U$ satisfies
\[
 |U\setminus V_i|\le|U|-|V_i|+|V_i\setminus U|
 \le(b+4/h)\tau|U|.
\]
It follows that $Q(s,R)\le(1+b+4/h)\tau$ for every start $s$ for which $U_s$ is defined. Averaging over the uniform start and using double stochasticity gives
$|R|/|C_0|\le\rho+(1+b+4/h)\tau$. Delete all edges leaving the selected sets $V_i$ and all edges incident to $R$. The number of changed neighbor entries is at most $2K_0\tau|C_0|+2d|R|$. This is $O_{d,h}((\rho+\tau)|C_0|)$, and the remaining nontrivial components expand by at least $h/4$. The sets and repair are used only in the soundness proof; the tester does not compute them.
\end{proof}

\subsection{Completeness, soundness, and query complexity}

\begin{proof}[Proof of Theorem~\ref{thm:main}]
Fix a sufficiently small absolute $\kappa$ as in Lemma~\ref{lem:cores}, and put $g=32\kappa\beta c^2/d^2$ and $h_*=dg/8$. Then $h_*\le\min\{d,c\}$ and $h_*/4=c_\beta$ as stated. Let $C_{\mathrm{comb}}\ge d$ and $\tau_0$ be supplied by Lemma~\ref{lem:combine} with $h=h_*$. In addition to the structural bounds on $a_0$, impose
\[
 C_1\eta\le\tau_0,\qquad
 C_{\mathrm{comb}}C_1\eta\le d\eps/4.
\]
It suffices to decrease $a_0$ further so that $a_0\le\min\{\tau_0/C_1,d/(4C_{\mathrm{comb}}C_1)\}$; this choice is independent of $n,\eps$.

Let $\rho$ be the fraction of all vertices that are not good. Apply Lemma~\ref{lem:combine} separately to every component, taking $Q=P^{2t}$ and $\tau=C_1\eta$. Summing the costs shows that deleting edges yields a graph in $\Xi_{c_\beta}(n)$ with at most
\[
 C_{\mathrm{comb}}(\rho+C_1\eta)n
\]
changed neighbor entries. Therefore an $\eps$-far graph must have
\[
 \rho\ge d\eps/(2C_{\mathrm{comb}}),
\]
since a smaller value would give repair cost less than $3d\eps n/4$.

Choose $C_{\mathrm{start}}$ in the tester of Section~\ref{sec:collisions} large enough that $N_{\mathrm{start}}=\lceil C_{\mathrm{start}}/\eps\rceil$ independent uniform starts miss all vertices at the displayed density with probability at most $1/6$. Proposition~\ref{prop:vertextest} and majority amplification ensure that a start violating Eq.~\eqref{eq:certificate} is rejected with probability at least $1-1/(100N_{\mathrm{start}})$, and a start whose component satisfies Eq.~\eqref{eq:mixed} is accepted with at least the same probability, for a sufficiently large absolute constant in the number of trials.

On a far graph, a vertex violating Eq.~\eqref{eq:certificate} is sampled with probability at least $5/6$. The amplified test rejects such a vertex except with probability at most $1/(100N_{\mathrm{start}})$, giving overall rejection probability greater than $2/3$.

For completeness, Lemma~\ref{lem:mixing} establishes Eq.~\eqref{eq:mixed} in every component of a graph in $\Xi_c(n)$. Proposition~\ref{prop:vertextest}, amplification, and a union bound give overall acceptance probability at least $99/100$.

The worst-case query bound is
\[
 \OO_{d,c,\beta}\left(\frac{\sqrt n}{\eps\eta^3}\right)
 =\OO_{d,c,\beta}\left(\frac{n^{1/2+\beta}}\eps+
                            \frac{\sqrt n}{\eps^4}\right),
\]
using $\eta=a_0\min\{\eps,n^{-\beta/3}\}$. All walk lengths and error-amplification factors are logarithmic in $n$ and $1/\eps$, with constants depending on $d,c,\beta$. All parameter bounds hold for every $n\ge1$.
\end{proof}

\begin{remark}[Fixed neighbor-table distance]
The distance in Section~\ref{sec:intro} minimizes over neighbor orderings of both graphs. If instead the input table's order is fixed, compacting a list after deletions can cause extra changes. Removing $r$ neighbors from one list can be implemented with at most $2r$ changed entries: replace deleted neighbors in the surviving prefix by surviving neighbors from the discarded suffix, then empty the suffix. Thus deleting $q$ edges costs at most $4q$ entries under the fixed-table convention, compared with $2q$ under our convention. Run the tester with proximity parameter $\eps/2$ to obtain soundness at $\eps$ under fixed-table distance. The repair proof uses deletions only, so this factor-two adjustment suffices; the displayed asymptotic query bound and $c_\beta$ are unchanged.
\end{remark}

\section{A lower bound on the expansion--complexity trade-off}\label{sec:tradeoff}

A sparse cut is sufficient to violate expansion, but it may be inexpensive to delete the edges crossing it. We first give a farness criterion that will be useful for both lower bounds. The idea is that a graph with positive expansion cannot be broken into small components without deleting many edges. If a repair instead leaves one large component, most of the original sparse cut remains inside it.

\begin{lemma}[Farness criterion from a sparse balanced cut]\label{lem:farness}
Fix $d$ and constants $h_0>0$, $c_*>h_1\ge0$, and $a\in(0,1/2]$. Let $G$ be a connected $N$-vertex graph of maximum degree at most $d$ with $h(G)\ge h_0$. Suppose there is $A\subseteq V(G)$ such that
\[
 aN\le |A|\le N/2,
 \qquad
 |E_G(A,V\setminus A)|\le h_1|A|.
\]
Then $G$ is $\eps_0$-far from $\Xi_{c_*}(N)$ for
\[
 \eps_0=\frac{a h_0(c_*-h_1)}{2d(h_0+c_*)}.
\]
\end{lemma}

\begin{proof}
Let $H\in\Xi_{c_*}(N)$ and let
\[
 M=|E(G)\setminus E(H)|+|E(H)\setminus E(G)|
\]
be the number of edges in the symmetric difference. If every connected component of $H$ has size at most $N/2$, then summing the $h_0$-expansion inequality in $G$ over the components of $H$ shows that at least $h_0N/2$ edges of $G$ run between different $H$-components. All of these edges are absent from $H$, so $M\ge h_0N/2$.

Otherwise $H$ has a unique component $C$ with $|C|>N/2$. Put $s=N-|C|<N/2$. Every $G$-edge between $C$ and $V\setminus C$ must be absent from $H$, and therefore
\begin{equation}\label{eq:farness-one}
 M\ge h_0s.
\end{equation}
If $s>|A|$, this already gives $M\ge h_0aN$. Assume henceforth that $s\le|A|$. Both sides of the cut of $C$ induced by $A$ have size at least $|A|-s$, since $|A\cap C|\ge|A|-s$ and $|C\setminus A|\ge N-s-|A|\ge|A|-s$. Since $H[C]$ has expansion at least $c_*$, it contains at least $c_*(|A|-s)$ edges crossing the cut $A,V\setminus A$. At most $h_1|A|$ of these can already be crossing edges of $G$; every remaining one belongs to $E(H)\setminus E(G)$. Hence
\begin{equation}\label{eq:farness-two}
 M\ge(c_*-h_1)|A|-c_*s.
\end{equation}
Combining Eqs.~\eqref{eq:farness-one} and~\eqref{eq:farness-two} and minimizing over $s\ge0$ gives
\[
 M\ge
 \frac{h_0(c_*-h_1)}{h_0+c_*}|A|
 \ge
 \frac{a h_0(c_*-h_1)}{h_0+c_*}N.
\]
The same lower bound on $M$ holds in the two earlier cases, since $a\le1/2$ and $(c_*-h_1)/(h_0+c_*)\le1$. A changed neighbor-table entry can remove one neighbor and add another, so it accounts for at most two edges in the symmetric difference. Hence the bounded-degree distance is at least the stated $\eps_0$.
\end{proof}

\subsection{Random $2$-lifts and noisy parities}

Fix $d$ and $c$ as in Theorem~\ref{thm:tradeoff}. Let $B$ be a uniformly random simple $d$-regular graph on $[N]$, where $dN$ is even. Given bits $Y_e$ on its edges, define the $2$-lift $G$ on $[N]\times\mathbb F_2$ by replacing every base edge $uv$ with the two edges
\[
 \{(u,i),(v,i+Y_{uv})\},\qquad i\in\mathbb F_2.
\]
The lift is simple and $d$-regular. In the YES distribution, the $Y_e$ are independent uniform bits. In the NO distribution, choose independent uniform bits $X_v$ on the base vertices and independent noise bits $Z_e\sim\operatorname{Ber}(p)$ on the base edges, and set $Y_{uv}=X_u+X_v+Z_{uv}$. Here $p>0$ will be a sufficiently small fixed constant. All additions of labels are in $\mathbb F_2$.

We recall the binary case of the noisy-parity indistinguishability result of Chiplunkar et al.~\cite{ChiplunkarKapralovKhannaMousavifarPeres2018}, using the formulation for simple regular graphs in Peng and Shen~\cite[Lemmas~6.1 and~6.9]{PengShen2026}. A source query specifies a base vertex and returns its entire ordered list of incident edges (including the matching half-edge indices) and their labels. For fixed $d$, there are constants $k_d,s_d>0$ such that, for $0<p\le1/4$, all sufficiently large admissible $N$, and any adaptive algorithm using at most $s_d N^{1/2+k_dp}$ source queries, the transcript distributions under uniform and noisy-parity labels have total variation distance less than $1/6$. Indeed, in Lemma~6.9 the binary noise has bias $1-2p$, and $-\log(1-2p)\ge2p$; its two conditions on the excess query exponent hold for $k_dp$ when $k_d$ is sufficiently small. A neighbor query to the lift is simulated by a single source query at the corresponding base vertex, using the returned edge label to determine the second coordinate. The simulation knows neither $X$ nor $Z$.

\subsection{Expansion of the lifts}

We use two properties of the random base graph. With probability $1-o(1)$, its adjacency eigenvalues other than the first (in decreasing order) have absolute value at most $2\sqrt{d-1}+o(1)$, by Friedman's theorem~\cite{Friedman2008}. Also, for a sufficiently small constant $\ell_d>0$, every ball of radius $\lfloor\ell_d\log N\rfloor$ contains at most one cycle. The latter follows from Bordenave's configuration-model estimate~\cite{Bordenave2020} and conditioning on simplicity; see Peng and Shen~\cite[Lemmas~6.2 and~6.3 and the proof of Lemma~6.5]{PengShen2026}. In particular, the base graph has expansion at least a fixed $h_B>0$ depending only on $d$.

Under uniform labels, the adjacency operator of the lift acts on functions that are equal on the two vertices of each fiber as the adjacency operator of $B$, and on functions that take opposite values as the signed adjacency operator of $B$ with independent uniform edge signs $(-1)^{Y_e}$. The random-signing theorem of Mohanty, O'Donnell, and Paredes~\cite[Theorem~3.1]{MohantyODonnellParedes2020}, as used in~\cite[Lemmas~6.4--6.5]{PengShen2026}, bounds the norm of this second operator by $2\sqrt{d-1}+o(1)$ with probability $1-o(1)$. Thus the second-largest adjacency eigenvalue of the lift is at most $2\sqrt{d-1}+o(1)$. The spectral lower bound on edge expansion~\cite{HooryLinialWigderson2006} gives $h(G)\ge(d-2\sqrt{d-1}-o(1))/2>c$ with probability $1-o(1)$. This uses the full lift, without deleting vertices from any fiber.

Under noisy-parity labels, relabeling $(v,i)$ as $(v,i+X_v)$ gives the lift with edge labels $Z_e$. We use these coordinates for the expansion analysis. Write $Z=\{e:Z_e=1\}$ for the noisy edges, so $|Z|=(1+o(1))pdN/2$ with probability $1-o(1)$ for fixed $p$. The layer $[N]\times\{0\}$ has $2|Z|$ boundary edges. The next lemma supplies the expansion lower bound needed for farness.

\begin{lemma}[Expansion under noisy labels]\label{lem:noisy-expansion}
For fixed $d\ge3$, there is a constant $p_0=p_0(d)>0$ such that, for every fixed $0<p\le p_0$, the NO lift satisfies $h(G)=\Omega_d(p)$ with probability $1-o(1)$ as $N\to\infty$ along admissible sizes.
\end{lemma}

\begin{proof}
Condition on a base graph satisfying the two properties above. Every nonempty connected set $C$ with $|C|\le N/2$ has at least two boundary edges. Indeed, if $|C|$ exceeds a sufficiently large constant depending on $h_B$, this follows from expansion. Otherwise $C$ lies in a ball containing at most one cycle for sufficiently large $N$, so $|E(B[C])|\le |C|$. Consequently $|\bd_B C|\ge(d-2)|C|$, which is at least two for $|C|\ge2$; a singleton has $d$ boundary edges. Here and below $\bd_B$ denotes the edge boundary in the base graph.

First consider a set in the lift containing exactly one vertex of each fiber. It is specified by a set $T\subseteq[N]$: choose the vertex with second coordinate $1$ above $T$ and the one with coordinate $0$ elsewhere. Its boundary has size $2|Z\triangle\bd_B T|$, since a base edge contributes two crossing edges exactly when its noise bit disagrees with the parity of the chosen coordinates. Put
\[
 W(Z)=\min_{T\subseteq[N]}|Z\triangle\bd_B T|.
\]
We show that $W(Z)=\Omega_d(pN)$ with high probability. Replacing $T$ by its complement preserves its boundary, so assume $|T|\le N/2$. The boundaries of the connected components of $B[T]$ are disjoint and partition $\bd_B T$. Therefore
\[
 W(Z)\ge |Z|-\sum_C (2|Z\cap\bd_B C|-|\bd_B C|)_+,
\]
where the sum ranges over all nonempty connected sets $C\subseteq[N]$ of size at most $N/2$, and $x_+=\max\{x,0\}$. A summand is positive only if strictly more than half the boundary edges are noisy.

For a fixed connected $C$ of size $k$, write $m_C=|\bd_B C|$. We have $\max\{2,h_B k\}\le m_C\le dk$. By independence of the noise bits,
\[
 \E[(2|Z\cap\bd_B C|-m_C)_+]
 \le dk\,2^{dk}p^{\max\{2,h_Bk/2\}}.
\]
The number of connected sets of size $k$ is at most $Nd^{2(k-1)}$: every such set is the set of vertices visited by a walk traversing a spanning tree in both directions. Choose a fixed integer $k_0$ with $h_Bk_0/2\ge2$. The sum of the expectation bounds over $k\le k_0$ is $O_d(p^2N)$. For $k>k_0$, use $p^{h_Bk/2}\le p^2(p^{h_B/2})^{k-k_0}$. Taking $p_0$ sufficiently small makes the remaining sum a convergent geometric series with a linear factor in $k$, also of order $O_d(p^2N)$. Thus, uniformly over the conditioned base graphs,
\[
 \E W(Z)\ge pdN/2-O_d(p^2N)=\Omega_d(pN)
\]
for sufficiently small $p_0$. Changing one noise bit changes $W(Z)$ by at most one, since each symmetric-difference size changes by at most one. The bounded-differences inequality gives $W(Z)=\Omega_d(pN)$ except with probability $\exp(-\Omega_d(p^2N))$. Every cut choosing one vertex per fiber therefore has $\Omega_d(pN)$ boundary edges.

Now let $S$ be an arbitrary nonempty subset of the lift with $|S|\le N$, and put $L_S=|\bd_G S|$. Let $F$ and $R$ be the sets of base vertices whose fibers contain, respectively, two and zero vertices of $S$. Every base edge crossing $F$ or crossing $R$ contributes at least one lifted edge to $\bd_G S$, so $|\bd_B F|,|\bd_B R|\le L_S$. Since $|F|\le N/2$, we have $|F|\le L_S/h_B$. If $|R|\ge N/2$, base expansion gives $N-|R|\le L_S/h_B$, hence $|S|\le2L_S/h_B$. Otherwise $|R|\le L_S/h_B$. In this case, removing one vertex from each full fiber and adding one to each empty fiber gives a set $S'$ containing exactly one vertex of each fiber. Its boundary satisfies
\[
 |\bd_G S'|\le L_S+d(|F|+|R|)\le (1+2d/h_B)L_S.
\]
The bound on cuts of this form implies $L_S=\Omega_d(pN)=\Omega_d(p|S|)$. In the first case $L_S\ge h_B|S|/2$, which also implies the claimed bound. Removing the conditioning on the base graph completes the proof.
\end{proof}

\subsection{Farness and indistinguishability}

\begin{proof}[Proof of Theorem~\ref{thm:tradeoff}]
Choose $p=k\gamma$ for a sufficiently small absolute constant $k>0$ and $\gamma_0$ small enough that $p\le p_0(d)$, $p\le1/4$, and $\gamma d<c$. With probability $1-o(1)$ the NO lift has expansion at least $h_0=\Omega_d(p)$ by Lemma~\ref{lem:noisy-expansion}, and its layer cut has expansion at most $2pd\le\gamma d/2$. Applying Lemma~\ref{lem:farness} with $c_*=\gamma d$, $h_1=\gamma d/2$, and $a=1/2$ yields distance at least $b\gamma$ from $\Xi_{\gamma d}(n)$ for some constant $b=b(d,c)>0$ independent of $\gamma$. This follows directly from the displayed distance bound, taking $h_0$ to be a fixed positive multiple of $p$. The YES lift lies in $\Xi_c(n)$ with probability $1-o(1)$.

Suppose a randomized tester has success probability at least $2/3$ on both promised classes and makes at most $s_d N^{1/2+k_dp}$ queries on every promised input. Truncate it at this budget on every input and simulate it by source queries. Its acceptance probabilities under the two lift distributions differ by at least $1/3-o(1)$, since the respective promises hold with probability $1-o(1)$, including for every $\eps\le b\gamma$. This contradicts the transcript distance bound of $1/6$ for sufficiently large $N$. Since $n=2N$ and $p=k\gamma$, the theorem follows with $a=k_dk$.
\end{proof}

The proximity regime in the theorem is natural for this construction. Replacing every noisy pair of crossing edges by the corresponding within-layer edges recovers two copies of the base graph, which has expansion at least $c$ with high probability. The cost is $O(pn)$ edge modifications for fixed $d$. Thus these NO instances are themselves $O_d(\gamma)$-close to $\Xi_c(n)$; the construction does not give a lower bound for proximity bounded away from zero as $\gamma$ tends to zero.

\section{A lower bound for Cayley graphs}\label{sec:lower}

We prove Theorem~\ref{thm:cayley-lower} using two graphs built on $L$ copies of the vertex set of the same bipartite LPS graph. The YES instance consists of $L$ disconnected copies of the base graph. In the NO instance, each chosen generator also advances a cyclic coordinate by one, and its inverse moves that coordinate back by one. A sequence of generators that returns to its starting vertex in the NO graph also returns to its starting vertex after projection to the LPS base.

Bencs et al. study self-intersections of one non-backtracking trajectory on regular high-girth expanders~\cite{BencsEtAl2026}. The oracle tester can instead branch and compare different explorations, so we must bound collisions in its entire adaptive transcript.

\subsection{The LPS base and a cyclic coordinate}

Fix a prime $p\equiv1\pmod4$ and put $d=p+1$ and $r=d/2$. The LPS construction gives an infinite family of bipartite $d$-regular Ramanujan Cayley graphs $X_q=\operatorname{Cay}(G_q,S_q)$, where $G_q=\operatorname{PGL}_2(\mathbb F_q)$ is the projective general linear group,\footnote{This is the group of invertible $2\times2$ matrices over $\mathbb F_q$, with nonzero scalar multiples identified.} $q\equiv1\pmod4$ is prime, $(p/q)=-1$, meaning that $p$ is a quadratic nonresidue modulo $q$, and $|G_q|=m_q=q(q^2-1)$~\cite{LubotzkyPhillipsSarnak1988}. We use the free-group description recalled by Alon and Bachmat~\cite{AlonBachmat2006}. There is a free group $F=\langle a_1,\ldots,a_r\rangle$\footnote{Its elements are words in the symbols $a_i^{\pm1}$ after cancelling adjacent inverse pairs. Multiplication concatenates two words and performs these cancellations.} and an onto map $\pi_q:F\to G_q$ satisfying $\pi_q(uv)=\pi_q(u)\pi_q(v)$. The symmetric generating set is
\[
 S_q=\{\pi_q(a_1)^{\pm1},\ldots,\pi_q(a_r)^{\pm1}\}.
\]
A word in $a_1^{\pm1},\ldots,a_r^{\pm1}$ is \emph{reduced} if it contains no adjacent inverse pair $a_i a_i^{-1}$ or $a_i^{-1}a_i$. Under the Cayley-graph interpretation, reduced words of length $k$ correspond exactly to $k$-step non-backtracking walks from a fixed vertex.

Write $N_q=\ker\pi_q$, the words whose image under $\pi_q$ is the identity. The graphs $X_q$ have girth at least $(4/3)\log_p m_q$ and all adjacency eigenvalues other than $\pm d$ have absolute value at most $2\sqrt p$~\cite{LubotzkyPhillipsSarnak1988,AlonBachmat2006}.

Fix an odd prime $L\ge5$ with $L\ne p$, whose size will be specified below. We restrict to an infinite subsequence of the primes $q$ above satisfying
\begin{equation}\label{eq:q-mod-L}
 L\nmid q(q^2-1).
\end{equation}
Such an infinite subsequence exists as follows. Choose a quadratic nonresidue $b_p$ modulo $p$ and a residue $b_L\not\equiv0,\pm1\pmod L$. By the Chinese remainder theorem choose a reduced residue class $q_0$ modulo $4pL$ with $q_0\equiv1\pmod4$, $q_0\equiv b_p\pmod p$, and $q_0\equiv b_L\pmod L$. Dirichlet's theorem gives infinitely many primes $q\equiv q_0\pmod{4pL}$; since $p\equiv1\pmod4$, quadratic reciprocity gives $(p/q)=(q/p)=-1$, while the choice modulo $L$ gives Eq.~\eqref{eq:q-mod-L}. Define $\phi:F\to\mathbb Z_L$ by $\phi(a_i)=1$, and let $K=\ker\phi$. Condition~\eqref{eq:q-mod-L} implies that $G_q$ has no quotient of order $L$. Therefore $\phi(N_q)=\mathbb Z_L$. Otherwise $N_q\subseteq K$, and $\phi$ would factor through $F/N_q\cong G_q$. Consequently the combined map $w\mapsto(\pi_q(w),\phi(w))$ from $F$ to $G_q\times\mathbb Z_L$ is surjective and has kernel $N_q\cap K$.

Set $n_q=Lm_q$. Define two $d$-regular Cayley graphs on $G_q\times\mathbb Z_L$:
\begin{align}
 Y_q&=\operatorname{Cay} \left(G_q\times\mathbb Z_L,
 \{(\pi_q(a_i),0)^{\pm1}:1\le i\le r\}\right),\label{eq:Yq}\\
 Z_q&=\operatorname{Cay} \left(G_q\times\mathbb Z_L,
 \{(\pi_q(a_i),1),(\pi_q(a_i)^{-1},-1):1\le i\le r\}\right).\label{eq:Zq}
\end{align}
Projection onto $G_q$ maps every walk in either graph to a walk in $X_q$. The graph $Y_q$ is the disjoint union of $L$ copies of $X_q$. In $Z_q$, the additional coordinate changes as specified in Eq.~\eqref{eq:Zq}. It is connected because the combined map $F\to G_q\times\mathbb Z_L$ is surjective, so the displayed set generates $G_q\times\mathbb Z_L$.

To prove farness later, we also need a positive lower bound on $h(Z_q)$ independent of $q$, though it may depend on the fixed parameter $L$. Such a bound will force any repair using few edits to leave almost all vertices in one component. We obtain it from expansion in the subgroup $K$ and the fixed quotient $F/K$, using the next lemma.

Throughout this subsection, the normalized adjacency spectral gap means
$1-\lambda_2$, where $\lambda_2$ is the second-largest eigenvalue of the
normalized adjacency operator.

\begin{lemma}[Spectral gap from a subgroup and a quotient]\label{lem:finite-index-gap}
Let $F=\langle S\rangle$ be generated by a finite symmetric set $S$, let $K$ be a normal subgroup of $F$ with $F/K$ finite,\footnote{A coset is a set $gK=\{gk:k\in K\}$. Normality means $gK=Kg$ for every $g\in F$; the cosets then form the quotient group $F/K$.} and let $H_j$ be normal subgroups of $F$ contained in $K$ with $F/H_j$ finite. Suppose there is a finite symmetric generating set $T\subseteq K$ and a constant $\kappa_K>0$ such that the normalized adjacency operator of every Cayley graph
\[
 \operatorname{Cay}(K/H_j,T)
\]
has spectral gap at least $\kappa_K$. Let $\overline S$ be the image of $S$ in $F/K$, retaining repetitions.\footnote{Different generators may have the same image in the quotient. Each original generator is still counted separately when averaging the walk's transitions.} Suppose also that the fixed quotient Cayley multigraph
\[
 \operatorname{Cay}(F/K,\overline S)
\]
has normalized adjacency spectral gap $\kappa_0>0$. If every $t\in T$ is a product of at most $\ell$ elements of $S$, then the normalized adjacency operators of the Cayley graphs $\operatorname{Cay}(F/H_j,S)$ have spectral gap at least
\[
 \min\left\{\kappa_0,\frac{\kappa_K}{|S|\ell^2}\right\}.
\]
\end{lemma}

\begin{proof}
Fix $j$ and let $\mathcal H$ be the space of complex-valued functions on $F/H_j$, with its usual inner product. For $g\in F/H_j$, let $(\pi(g)v)(x)=v(xg)$. These maps permute the coordinates, so they preserve the inner product, and $\pi(gh)=\pi(g)\pi(h)$. For a finite symmetric set $R$, write $\mathcal E_R(v):=(2|R|)^{-1}\sum_{r\in R}\|\pi(r)v-v\|^2$. Let $\mathcal H^K=\{v:\pi(k)v=v\text{ for every }k\in K\}$. This subspace consists of functions constant on each coset of $K/H_j$. It and its orthogonal complement are preserved by every $\pi(g)$ because $K$ is normal in $F$. On $\mathcal H^K$, the maps $\pi(g)$ act as translations on the quotient $F/K$. Hence, on the part of $\mathcal H^K$ orthogonal to the constant functions, $\mathcal E_S(v)\ge\kappa_0\|v\|^2$.

Now let $v\perp\mathcal H^K$. Translation by elements of $T$ acts separately on each coset of $K/H_j$, and on each coset it gives the walk on $\operatorname{Cay}(K/H_j,T)$. By the assumed spectral gap, $\mathcal E_T(v)\ge\kappa_K\|v\|^2$. For $t=s_1\cdots s_m$ with $m\le\ell$ and $s_i\in S$, telescoping and Cauchy--Schwarz give
\begin{align*}
 \|\pi(t)v-v\|^2
 &\le m\sum_{i=1}^m\|\pi(s_i)v-v\|^2\\
 &\le \ell^2\sum_{s\in S}\|\pi(s)v-v\|^2.
\end{align*}
Averaging over $t\in T$ yields $\mathcal E_T(v)\le|S|\ell^2\mathcal E_S(v)$, and therefore $\mathcal E_S(v)\ge\kappa_K\|v\|^2/(|S|\ell^2)$. The two invariant subspaces are orthogonal and preserved by the normalized $S$-averaging operator. The preceding estimates therefore bound the spectral gap on functions orthogonal to the constants, as required.
\end{proof}

\begin{lemma}[Expansion properties of $Y_q$ and $Z_q$]\label{lem:YN-geometry}
There are constants $c_+=c_+(d)>0$ and $h_0=h_0(d,L)>0$ such that, for every $q$ in the chosen subsequence,
\begin{enumerate}[label=(\roman*),leftmargin=2.4em]
    \item $Y_q\in\Xi_{c_+}(n_q)$;
    \item $h(Z_q)\ge h_0$; and
    \item $Z_q$ has a set $A_q$ with $|A_q|=\lfloor L/2\rfloor m_q$ and
    \[
      \frac{|E(A_q,V\setminus A_q)|}{|A_q|}
      =\frac{d}{\lfloor L/2\rfloor}.
    \]
\end{enumerate}
\end{lemma}

\begin{proof}
Since $X_q$ is $d$-regular Ramanujan, the standard Cheeger inequality gives a uniform positive lower bound on its edge expansion~\cite{HooryLinialWigderson2006}. Fix any $0<c_+<(d-2\sqrt{d-1})/2$. Every component of $Y_q$ is a copy of $X_q$, proving (i).

For (ii), the LPS quotients $F/N_q$ form an expander family~\cite{LubotzkyPhillipsSarnak1988}. Ab\'ert and Elek~\cite[Theorem~8]{AbertElek2012} show that an expanding family of finite quotients by normal subgroups remains expanding after restriction to a fixed finite-index subgroup, for any fixed finite symmetric generating set of that subgroup. Here each $N_q$ is normal in $F$, and $K$ has fixed index $L$ in $F$. Thus, for a fixed finite symmetric generating set $T$ of $K$, the Cayley graphs $\operatorname{Cay}(K/(N_q\cap K),T)$ form an expander family. Since $T$ is fixed, Cheeger's inequality gives a constant $\kappa_K>0$, independent of $q$, bounding their normalized adjacency spectral gaps from below.

Every element of $T$ can be written as a product of at most $\ell=\ell(L)$ elements of $S=\{a_1^{\pm1},\ldots,a_r^{\pm1}\}$. The quotient $F/K\cong\mathbb Z_L$ is fixed. Retaining multiplicities in the image of $S$, each $a_i$ contributes $+1$ and each $a_i^{-1}$ contributes $-1$, so the normalized quotient adjacency operator has nontrivial eigenvalues $\cos(2\pi j/L)$ for $1\le j\le L-1$. Its spectral gap is therefore $\kappa_0=1-\cos(2\pi/L)>0$. Lemma~\ref{lem:finite-index-gap}, applied with $H_j=N_q\cap K$, gives a uniform normalized spectral gap for the Cayley graphs $Z_q$. Cheeger's inequality then gives $h(Z_q)\ge h_0(d,L)>0$.

For (iii), let $I\subseteq\mathbb Z_L$ consist of $k=\lfloor L/2\rfloor$ consecutive residues and set $A_q=G_q\times I$. Across each of the two edges leaving $I$ in the cycle $\mathbb Z_L$, the $r=d/2$ generators indexed by $a_1,\ldots,a_r$ contribute exactly $rm_q$ undirected edges. Thus $|E(A_q,V\setminus A_q)|=2rm_q=dm_q$, which gives the claimed ratio.
\end{proof}

\subsection{Constant distance from the target property}

Lemma~\ref{lem:YN-geometry} gives the two ingredients of the farness criterion in Lemma~\ref{lem:farness}: a positive expansion lower bound and a balanced cut of small expansion.

Choose the fixed prime $L$ large enough that $c_-:=2d/\lfloor L/2\rfloor<c_+$. Then Lemma~\ref{lem:YN-geometry} (iii) has $h_1=c_-/2$. Applying Lemma~\ref{lem:farness} with $a=\lfloor L/2\rfloor/L$ and $c_*=c_-$ gives a constant $\eps_0>0$ such that every $Z_q$ is $\eps_0$-far from $\Xi_{c_-}(n_q)$. Since $L$ is fixed before $q$ varies, all constants remain independent of $n_q$.

\subsection{Return probabilities in the LPS graph}

Let $R_q(k)$ be the probability that a uniformly random reduced word of length $k$ in the free generators $a_1^{\pm1},\ldots,a_r^{\pm1}$ belongs to $N_q$. Equivalently, $R_q(k)$ is the return probability of the length-$k$ non-backtracking walk on $X_q$. There are $dp^{k-1}$ reduced words of length $k$.

At intermediate lengths, the arithmetic description of LPS walks reduces the return bound to counting representations of an integer $N$ as a sum of three squares. We need a uniform $O(\sqrt N\log(2N))$ bound; its logarithmic factor will account for the logarithmic loss in the query lower bound. Earlier cycle-count estimates for these Ramanujan constructions were obtained by Tillich and Z\'emor~\cite{TillichZemor1997}. We use the counting argument of Alon and Bachmat~\cite{AlonBachmat2006}, with the following uniform three-squares bound.

\begin{lemma}[A uniform three-squares bound]\label{lem:three-squares}
Let
\[
 r_3(N)=\bigl|\{(x_1,x_2,x_3)\in\mathbb Z^3:
 x_1^2+x_2^2+x_3^2=N\}\bigr|.
\]
There is an absolute constant $C$ such that, for every integer $N\ge1$,
\[
 r_3(N)\le C\sqrt N\,\log(2N).
\]
\end{lemma}

\begin{proof}
The case $N=1$ is immediate, so assume $N\ge2$. Repeatedly using the classical identity $r_3(4N)=r_3(N)$ reduces to the case $4\nmid N$. Gauss's three-squares formula expresses $r_3(N)$ as a constant multiple of a quantity $H(M)$ with $M\le4N$;\footnote{The quantity $H(M)$ is called a Hurwitz class number. The stated formula for $H(M)$ is the only class-number fact used in the estimate.} see Mortenson~\cite{Mortenson2017} for a modern proof of the classical formula and Filaseta, Klein, and Sabuncu~\cite{FilasetaKleinSabuncu2026} for a convenient class-number formulation. So it is enough to prove
\begin{equation}\label{eq:Hurwitz-upper}
 H(M)=O(\sqrt M\log(2M)).
\end{equation}

If $H(M)=0$, there is nothing to prove. Otherwise the class-number formula (see~\cite{FilasetaKleinSabuncu2026}) gives integers $D<0$ and $f\ge1$ with $-M=Df^2$, and a function $\chi_D:\mathbb Z\to\{-1,0,1\}$,\footnote{In the cited class-number formula, $\chi_D$ is the quadratic Dirichlet character associated with $D$.} such that
\[
 H(M)
 =
 \frac{\sqrt{|D|}}{\pi}L(1,\chi_D)
 \sum_{e\mid f}\mu(e)\chi_D(e)\sigma(f/e),
\]
Here $L(1,\chi_D)=\sum_{m\ge1}\chi_D(m)/m$, $\mu$ is the M\"obius function, and $\sigma$ is the sum-of-divisors function.\footnote{Here $\mu(n)=0$ if a prime square divides $n$, and $\mu(n)=(-1)^r$ if $n$ is a product of $r$ distinct primes; $\mu(1)=1$. Also $\sigma(n)=\sum_{a\mid n}a$ and $\varphi(n)=|\{1\le a\le n:\gcd(a,n)=1\}|$.} The function satisfies $\chi_D(ab)=\chi_D(a)\chi_D(b)$, is periodic modulo $|D|$, vanishes on integers not coprime to $|D|$, and has sum zero over one period. We also write $\varphi$ for Euler's totient function and $\operatorname{rad}(f)$ for the product of the distinct prime divisors of $f$. Put $S_D(f):=\sum_{e\mid f}\mu(e)\chi_D(e)\sigma(f/e)$. For coprime $u,v$, we have $S_D(uv)=S_D(u)S_D(v)$. If $\ell$ is prime and $\ell^j\mid f$ but $\ell^{j+1}\nmid f$, then
\[
 S_D(\ell^j)
 =\sigma(\ell^j)-\chi_D(\ell)\sigma(\ell^{j-1}).
\]
To retain only terms coprime to $f$ in the series, set
\[
 L_f(1,\chi_D)
 :=L(1,\chi_D)\prod_{\ell\mid f}\left(1-\frac{\chi_D(\ell)}{\ell}\right).
\]
A direct check of the cases $\chi_D(\ell)\in\{-1,0,1\}$ gives
\[
 \frac{|S_D(\ell^j)|}{\ell^j}
 \left|1-\frac{\chi_D(\ell)}{\ell}\right|^{-1}
 \le\frac{\ell}{\ell-1}.
\]
Multiplying over $\ell\mid f$ yields
\begin{equation}\label{eq:Hurwitz-local}
 |L(1,\chi_D)S_D(f)|
 \le f\,|L_f(1,\chi_D)|\,\frac{f}{\varphi(f)}.
\end{equation}

Bounding $L_f(1,\chi_D)$ and $f/\varphi(f)$ separately would introduce an unnecessary $\log\log f$ loss. Instead we retain the factor $\varphi(f)/f$ in the bound for $L_f(1,\chi_D)$; Eq.~\eqref{eq:Lf-density} below then cancels the factor $f/\varphi(f)$ in Eq.~\eqref{eq:Hurwitz-local}. Expand
\[
 L_f(1,\chi_D)=\sum_{m\ge1}\frac{a_m}{m},
 \qquad
 a_m=\chi_D(m)\mathbf1_{(m,f)=1}.
\]
The coefficients are periodic with period $Q_{\mathrm{per}}=\operatorname{lcm}(|D|,\operatorname{rad}(f))$, and their sum over one period is zero. Indeed, the coprimality condition is redundant at primes dividing $D$, while at the remaining primes it factors independently by the Chinese remainder theorem; $\chi_D$ has zero sum over a complete period.

For the first period,
\begin{align*}
 \sum_{m\le Q_{\mathrm{per}}}\frac{|a_m|}{m}
 &\le\sum_{\substack{m\le Q_{\mathrm{per}}\\(m,f)=1}}\frac1m\\
 &\le \prod_{\substack{\ell\le Q_{\mathrm{per}}\\\ell\nmid f}}\left(1-\frac1{\ell}\right)^{-1}\\
 &=\prod_{\ell\le Q_{\mathrm{per}}}\left(1-\frac1{\ell}\right)^{-1}
   \prod_{\ell\mid f}\left(1-\frac1{\ell}\right)\\
 &=O  \left(\frac{\varphi(f)}{f}\log(2Q_{\mathrm{per}})\right),
\end{align*}
where we used Mertens' bound $\prod_{\ell\le x,\,\ell\ \mathrm{prime}}(1-1/\ell)^{-1}=O(\log(2x))$. Moreover, the number of nonzero coefficients in one period is at most $\varphi(Q_{\mathrm{per}})\le Q_{\mathrm{per}}\varphi(f)/f$. Since the complete-period sum is zero, every partial sum of $(a_m)$ is therefore $O(Q_{\mathrm{per}}\varphi(f)/f)$. Summation by parts on the tail gives $|\sum_{m>Q_{\mathrm{per}}}a_m/m|=O(\varphi(f)/f)$. Consequently
\begin{equation}\label{eq:Lf-density}
 |L_f(1,\chi_D)|
 =O  \left(\frac{\varphi(f)}f\log(2Q_{\mathrm{per}})\right)
 =O  \left(\frac{\varphi(f)}f\log(2M)\right).
\end{equation}
Substituting Eq.~\eqref{eq:Lf-density} into Eq.~\eqref{eq:Hurwitz-local} cancels the factor $f/\varphi(f)$ and gives $|L(1,\chi_D)S_D(f)|=O(f\log(2M))$. Since $\sqrt{|D|}f=\sqrt M$, Eq.~\eqref{eq:Hurwitz-upper} follows, and hence so does the lemma.
\end{proof}

\begin{lemma}[Uniform non-backtracking return bound]\label{lem:LPS-return}
There is a constant $C=C(p)<\infty$ such that, for all sufficiently large $q$ in the chosen subsequence and every $k\ge1$,
\[
 R_q(k)\le C\frac{\log(3m_q)}{m_q}.
\]
\end{lemma}

\begin{proof}
For $k<\operatorname{girth}(X_q)$ the return probability is zero. Suppose $\operatorname{girth}(X_q)\le k\le2\log_p m_q$. The arithmetic description of the LPS construction~\cite{LubotzkyPhillipsSarnak1988,AlonBachmat2006} identifies closed non-backtracking walks of length $k$ with normalized integer solutions of
\begin{equation}\label{eq:LPS-norm}
 x_0^2+q^2(x_1^2+x_2^2+x_3^2)=p^k.
\end{equation}
Here a solution is normalized if $x_0>0$ is odd, $x_1,x_2,x_3$ are even, and $p$ does not divide all four coordinates.\footnote{The last condition excludes $(p^{k/2},0,0,0)$ for positive even $k$. Without it, the equation would have solutions even at lengths below the graph's girth.} The possible values of $x_0$ lie in the union of two arithmetic progressions of spacing $2q^2$~\cite[Theorem~3.4]{LubotzkyPhillipsSarnak1988}; see also~\cite[Theorem~3.1]{AlonBachmat2006}, which records the same statement in the non-backtracking-walk formulation used here. Since $|x_0|\le p^{k/2}$, the two progressions contain $O(1+p^{k/2}/q^2)$ possible values in this range. The same LPS arithmetic statement implies $p^{k/2}>q^2$ whenever such a normalized solution exists~\cite[Theorem~3.1]{AlonBachmat2006}, so this is $O(p^{k/2}/q^2)$. For a fixed admissible $x_0$, put $M=(p^k-x_0^2)/q^2$. A normalized solution has $M>0$, and $M\le p^k/q^2$. Lemma~\ref{lem:three-squares} bounds the number of possible triples $(x_1,x_2,x_3)$ by $O(p^{k/2}q^{-1}\log(3m_q))$, where we used $k\le2\log_p m_q$. Hence the number of normalized solutions to Eq.~\eqref{eq:LPS-norm}, and therefore the number of closed non-backtracking walks of length $k$ from a fixed vertex, is at most $O(p^kq^{-3}\log(3m_q))$. Dividing by the $dp^{k-1}=\Theta_p(p^k)$ reduced words of length $k$ gives
\[
 R_q(k)=O \left(\frac{\log(3m_q)}{q^3}\right)
 =O \left(\frac{\log(3m_q)}{m_q}\right).
\]

For longer walks we use the Ramanujan eigenvalue bound and the standard recurrence for non-backtracking walks on regular graphs~\cite{AlonBenjaminiLubetzkySodin2007}. Let $Q_k$ be the non-backtracking polynomial, so that $Q_k(A_{X_q})(v,v)$ counts closed non-backtracking walks of length $k$ from $v$. It satisfies
\[
 Q_0=1,\quad Q_1=z,\quad Q_2=z^2-d,
 \qquad
 Q_k=zQ_{k-1}-pQ_{k-2}\quad(k\ge3).
\]
For $|z|\le2\sqrt p$, this recurrence gives $|Q_k(z)|\le C_p(k+1)p^{k/2}$.\footnote{Write $z=2\sqrt p\cos\vartheta$. For $k\ge2$, the recurrence gives $Q_k(z)=p^{k/2}\bigl(\sin((k+1)\vartheta)-p^{-1}\sin((k-1)\vartheta)\bigr)/\sin\vartheta$. Use $|\sin(j\vartheta)|\le j|\sin\vartheta|$, taking limits at $\vartheta=0,\pi$; the case $k=1$ is immediate.} The graph $X_q$ is bipartite and Ramanujan, so its only eigenvalues outside this interval are $\pm d$. Left multiplication by a group element preserves adjacency and can map any vertex to any other. Thus every diagonal entry of $Q_k(A_{X_q})$ equals $m_q^{-1}\operatorname{tr}(Q_k(A_{X_q}))$. Dividing by the total number $dp^{k-1}$ of reduced words gives $R_q(k)\le2/m_q+C_pk p^{-k/2}$. For $k\ge2\log_p m_q$, the second term is $O((\log m_q)/m_q)$ at the left endpoint and decreases thereafter. This proves the claimed bound for all sufficiently large $q$.
\end{proof}

Every reduced word that represents the identity in $Z_q$ lies in $N_q\cap K\subseteq N_q$. Hence the non-backtracking return probability of $Z_q$ is pointwise bounded by $R_q(k)$.

\subsection{Coupling adaptive query transcripts}

We use the following coupling to compare adaptive query transcripts. Consider a Cayley graph $G=\operatorname{Cay}(\Gamma,\Sigma)$ with a fixed symmetric generator indexing $\Sigma=\{s_1^{\pm1},\ldots,s_r^{\pm1}\}$, where $d=2r$. For $k\ge1$, let $R_G(k)$ be the probability that a uniformly random reduced word of length $k$ in these indexed generators evaluates to the identity of $\Gamma$. Randomizing the adjacency-list order at every group element hides this generator indexing from the oracle algorithm.

The proof below avoids conditioning on the absence of earlier collisions. We first run the algorithm against an ideal $d$-regular forest with independent random neighbor-list orderings and then, separately for each Cayley graph, assign hidden generator labels to the forest and project it to the finite group. The word along any fixed path in this ideal forest is uniformly distributed over reduced words of the corresponding length, without conditioning on earlier collisions, even though the algorithm is adaptive.

\begin{lemma}[Coupling to an ideal forest]\label{lem:transcript}
Let $U=\operatorname{Cay}(\Gamma_U,\Sigma_U)$ and
$V=\operatorname{Cay}(\Gamma_V,\Sigma_V)$ be simple $d$-regular
Cayley graphs with fixed symmetric generator indexings on $n$ vertices.
Suppose that $R_U(k),R_V(k)\le R(k)$ for every $k\ge1$.
There is a constant $C_d>0$, depending only on $d$, such that for every
integer $Q\ge1$, under independent random vertex relabelings and independent
random adjacency-list orderings, the transcript distributions of every
deterministic adaptive algorithm making at most $Q$ oracle queries satisfy
\[
 d_{\mathrm{TV}}(\mathsf T_U,\mathsf T_V)
 \le C_d Q^2\left(\frac1n+\sup_{1\le k\le2Q}R(k)\right).
\]
The algorithm may query arbitrary labels in $[n]$, including labels
that were not returned by previous oracle answers.
\end{lemma}

\begin{proof}
It is enough to compare the transcript for either graph
$G=\operatorname{Cay}(\Gamma,\Sigma)$ with one graph-independent
ideal transcript. If $Q>n/4$, the displayed bound is trivial after
increasing $C_d$, so assume $Q\le n/4$.

Define an ideal oracle $\mathcal O_\infty$ that maintains a forest of
copies of the infinite $d$-regular tree. Its answers use labels in $[n]$,
which we distinguish from the vertices of the forest. At each forest
vertex, assign the $d$ incident tree edges to the adjacency-list positions
in $[d]$ by an independent uniformly random bijection, revealed only as
queries require. When the algorithm queries a previously unseen vertex
label, that label starts a new root. When it queries a previously
unqueried adjacency-list position of a vertex already in the forest,
the oracle follows the corresponding tree edge. If the neighboring forest
vertex has not appeared before, it receives a uniformly random unused
vertex label. Otherwise, the oracle returns its existing label. Queries
to a previously queried position return the same answer as before.
In particular, querying the position leading back along a revealed tree
edge returns the parent's label and is not a collision. Let
$\mathsf T_\infty$ be this ideal transcript. Its distribution depends
only on $d$ and the deterministic algorithm.

We couple the randomized oracle for $G$ to $\mathcal O_\infty$. Assign
each new ideal root to a uniformly random group element not already
assigned to a vertex label. Independently for each tree in the ideal
forest, assign hidden generators to its incident edges as follows. At a
root, use a uniformly random bijection between its $d$ incident edges
and $\Sigma$. If a forest vertex is reached from its parent along a
generator $s$, force the parent edge at the child to carry $s^{-1}$ and
assign the remaining $d-1$ generators uniformly to the remaining child
edges. The image of any forest vertex is obtained by multiplying the
root image by the word on the unique root-to-vertex path; write $\psi(v)$
for the resulting group image of a forest vertex $v$.

Different trees in the ideal forest are allowed to use independent local
generator assignments even if they eventually acquire the same group
image. Thus the auxiliary construction need not define one global
adjacency-list ordering after such an identification. Before the first
identification, however, all finite group vertices encountered are
distinct, and the independent local assignments are exactly a uniformly
random collection of adjacency-list orderings at those vertices.
Coupling the unused vertex labels through the uniform random vertex
relabeling of $G$, the actual transcript and $\mathsf T_\infty$ are
therefore identical unless two distinct forest vertices receive the
same group image.

After $Q$ queries the ideal forest contains at most $2Q$ forest vertices, because one query can introduce at most one new root and at
most one new returned vertex. Condition on the ideal transcript and its revealed forest. The hidden generator assignments are independent of this transcript, so they retain their original distribution. Fix two distinct vertices $a,b$ in the revealed forest; the probabilities below are conditional on this ideal transcript.

Suppose first that they lie in the same tree, at tree distance $k$.
Then $1\le k\le2Q$. Along the unique path from $a$ to $b$, the reverse of the edge just
traversed carries the inverse generator. That edge is excluded by
non-backtracking, and every other outgoing generator is chosen uniformly. Consequently the hidden word on this path
is exactly uniform among the $d(d-1)^{k-1}$ reduced words of length $k$.
Since $\psi(a)=\psi(b)$ precisely when this word evaluates to the identity,
\[
 \Pr[\psi(a)=\psi(b)]=R_G(k)\le R(k).
\]
This equality holds without conditioning on the absence of earlier
collisions.

Now suppose that $a$ and $b$ lie in different trees. Without loss of
generality, let $b$ lie in the tree whose root was created later. Condition
on the entire coupling history up to, but not including, the group image
assigned to that later root. Also condition on
the hidden generator assignments along the two root-to-vertex paths; these
quantities are independent of the later root image. At the time the later
root was created, its image was uniform among the group elements not yet
assigned to visible vertex labels. At least $n-2Q$ group elements were
available.

Once the earlier root image and the two path words are fixed, the equality
$\psi(a)=\psi(b)$ determines at most one possible value of the later root
image. Therefore
\[
 \Pr[\psi(a)=\psi(b)]
 \le \frac1{n-2Q}
 \le \frac2n,
\]
where the last inequality uses $Q\le n/4$. This also covers trees started
by arbitrary vertex labels chosen adaptively by the algorithm.

There are at most $\binom{2Q}{2}=O(Q^2)$ pairs of distinct forest vertices. A union bound gives
\[
 \Pr[\psi\text{ is not injective}]
 \le C_dQ^2\left(\frac1n+\sup_{1\le k\le2Q}R(k)\right).
\]
The bound is uniform in the conditioned ideal transcript. Averaging over it and using the coupling inequality gives
\[
 d_{\mathrm{TV}}(\mathsf T_G,\mathsf T_\infty)
 \le C_dQ^2\left(\frac1n+\sup_{1\le k\le2Q}R(k)\right).
\]
Applying this estimate to $U$ and to $V$ and using the triangle inequality
proves the lemma, after absorbing a factor of two into $C_d$.
\end{proof}

For $Y_q$, a reduced word in the indexed generators evaluates to the identity exactly when it belongs to $N_q$, so $R_{Y_q}(k)=R_q(k)$. For $Z_q$, a reduced word evaluates to the identity only when it belongs to $N_q\cap K\subseteq N_q$, and hence $R_{Z_q}(k)\le R_q(k)$. Lemma~\ref{lem:transcript} therefore applies to the hard pair with $R(k)=R_q(k)$.

\subsection{Proof of the lower bound}

\begin{proof}[Proof of Theorem~\ref{thm:cayley-lower}]
Fix $p$, then choose $L$, $c_+$, $c_-$, and $\eps_0$ as above, and let $\mathcal N=\{n_q=Lq(q^2-1):q\text{ lies in the chosen subsequence}\}$. For each $n_q$, let $\mathcal Y_q$ and $\mathcal Z_q$ be the distributions over bounded-degree oracle representations obtained by independently randomizing the vertex names and adjacency-list orderings of $Y_q$ and $Z_q$. The underlying graph of every representation in the support of $\mathcal Y_q$ lies in $\Xi_{c_+}(n_q)$, while the underlying graph of every representation in the support of $\mathcal Z_q$ is a Cayley graph that is $\eps_0$-far from $\Xi_{c_-}(n_q)$.

Since $L$ is constant, $m_q=\Theta(n_q)$. Let $\mathsf T_{\mathcal Y_q}$ and $\mathsf T_{\mathcal Z_q}$ denote the resulting transcript distributions of a fixed deterministic algorithm under these two randomized oracle representations. Lemmas~\ref{lem:LPS-return} and~\ref{lem:transcript} give, for any deterministic $Q$-query algorithm,
\[
 d_{\mathrm{TV}}(\mathsf T_{\mathcal Y_q},\mathsf T_{\mathcal Z_q})
 \le C Q^2\frac{\log(3n_q)}{n_q}
\]
for all sufficiently large $q$, after increasing $C$ to absorb the $1/n_q$ term.

Choose a sufficiently small constant $a=a(d,L)>0$. If $Q\le a\sqrt{n_q/\log(3n_q)}$, then the transcript distance is less than $1/3$ for all sufficiently large $q$. Such a deterministic algorithm cannot distinguish $\mathcal Y_q$ from $\mathcal Z_q$ with success probability $2/3$. Yao's minimax principle~\cite{Yao1977} therefore gives the same lower bound for randomized algorithms. Since replacing $\log(3n)$ by $\log n$ changes only an absolute constant for sufficiently large $n$, the theorem follows.
\end{proof}

\section{Limitations and open questions}\label{sec:open}

For fixed $d,c$ in the range of Theorem~\ref{thm:tradeoff}, our bounds determine the dependence of the excess query exponent on the rejection expansion threshold $\gamma d$ up to constant factors when $\eps\le b\gamma$ and all parameters are fixed. The dependence on the rejection threshold when $\eps$ is fixed independently of $\gamma$ remains open. In particular, the lower bound does not rule out a near-square-root tester with a positive rejection threshold that depends on $\eps$. The quantitative dependence on $c$ is also open. In the upper bound, the tester pays for $O(1/\eta)$ further return estimates at each start, and it is unclear whether this cost is necessary.

For the lower bound, the remaining quantitative question is whether the Cayley-graph promise still permits an $\Omega(\sqrt n)$ bound. Our $\sqrt{\log n}$ loss comes from the intermediate-length relation estimate in Lemma~\ref{lem:LPS-return}. Since a uniform $O(\sqrt N)$ bound for the number $r_3(N)$ of representations as a sum of three squares is false in general~\cite{FilasetaKleinSabuncu2026}, removing this loss would require additional information specific to the LPS relations, a sharper adaptive transcript argument, or a different hard family. Our construction also leaves open a lower bound of this order at every sufficiently large input size.

\bibliographystyle{plain}
\bibliography{refs}
\appendix

\section{A simpler tester for walk-regular inputs}\label{app:walkregular}

When return probabilities are constant within each component, comparing the two lengths already supplies the local conditions. The tester can then omit the further walks from collision locations. We record the resulting improvement in the dependence on the comparison accuracy.

A graph $H$ with adjacency matrix $A_H$ is \emph{walk-regular} if $(A_H^k)_{v,v}$ is independent of $v$ for every integer $k\ge0$~\cite{GodsilMcKay1980}. A graph is \emph{componentwise walk-regular} if each component is walk-regular; different components need not have the same return probabilities. Taking $k=2$ shows that every walk-regular graph is regular. On a component of degree $r$, the lazy walk matrix is $(1-r/(2d))I+A_H/(2d)$. Its powers are polynomials in $A_H$, so their diagonal entries are constant as well. If the whole graph is walk-regular, the diagonal entries are constant across all vertices.

\begin{proposition}\label{prop:restricted-tester}
For the parameters of Theorem~\ref{thm:main}, there is a pseudo-tester accepting every graph in $\Xi_c(n)$ and rejecting every componentwise walk-regular graph $\eps$-far from $\Xi_{c_\beta}(n)$, each with probability at least $2/3$, using
\[
 \widetilde O_{d,c,\beta}\!\left(
       \frac{n^{1/2+\beta/3}}{\eps}+\frac{\sqrt n}{\eps^2}\right)
\]
queries. If the whole input is walk-regular, one starting vertex suffices, and the bound becomes
\[
 \widetilde O_{d,c,\beta}\!\left(n^{1/2+\beta/3}+\frac{\sqrt n}{\eps}\right).
\]
\end{proposition}

We first prove the collision comparison used by this tester. For a component $C$, put
\[
 M_C=\max_{x\in C}A_x,\qquad b_C=\min_{x\in C}B_x,
 \qquad\Delta_C=M_C/b_C-1.
\]
Equations~\eqref{eq:return-range} and~\eqref{eq:cs} imply
\begin{equation}\label{eq:return-sandwich}
 b_C\le B_x\le A_x\le(1+\Delta_C)b_C\qquad(x\in C)
\end{equation}
and
\begin{equation}\label{eq:appendix-pointwise}
 \|P^{2t}(x,\cdot)\|_\infty\le M_C.
\end{equation}
For a distribution $r$, write $F_j(r)=\sum_vr(v)^j$. From $q\ge2$ independent samples, let $\widehat F_2(r)$ be the number of pairwise collisions divided by $\binom q2$. Lemma~\ref{lem:statistics}, taking $D$ to be the whole domain, gives
\begin{equation}\label{eq:refined-variance}
 \E\widehat F_2(r)=F_2(r),\qquad
 \Var(\widehat F_2(r))\le C\left(\frac{F_2(r)}{q^2}
             +\frac{F_3(r)-F_2(r)^2}{q}\right)
\end{equation}
for an absolute constant $C$. The difference $F_3(r)-F_2(r)^2$ is the variance of $r(X)$ for $X$ drawn from $r$. In particular, it vanishes on a uniform distribution. The relation between the two walk lengths gives the following bounds.

\begin{lemma}[Third moments of the two walk distributions]\label{lem:walk-third-moment}
Fix a component $C$, put $b=b_C$ and $\delta=\Delta_C$, and suppose $0\le\delta\le1$. For any $x\in C$, let $r=P^t(x,\cdot)$ and $p=P^{2t}(x,\cdot)$. Then
\[
 F_3(p)-F_2(p)^2=O(\delta b^2),
 \qquad
 F_3(r)-F_2(r)^2=O(\delta b^2+\delta^{3/2}b^{3/2}),
\]
with absolute implicit constants.
\end{lemma}

\begin{proof}
Put $a=\|r\|_2^2$, $b_x=\|p\|_2^2$, and $D=a-b_x$. Equation~\eqref{eq:return-sandwich} gives
\[
 b\le b_x\le a\le(1+\delta)b,\qquad 0\le D\le\delta b.
\]
Equation~\eqref{eq:appendix-pointwise} gives $\|p\|_\infty\le(1+\delta)b=:M$. Thus, with $G_p=F_3(p)-b_x^2$,
\[
 0\le G_p\le(M-b_x)b_x\le\delta(1+\delta)b^2.
\]
This proves the first bound.

To control the short walk, set $e=r-p$. Spectral decomposition of $P$ gives
\begin{equation}\label{eq:semigroup-difference}
 \|e\|_2^2\le D,
\end{equation}
because $\lambda^{2t}(1-\lambda^t)^2\le\lambda^{2t}(1-\lambda^{2t})$ for $0\le\lambda\le1$. Also $\langle p,e\rangle=(D-\|e\|_2^2)/2\le D/2$. Consequently,
\begin{align*}
 \sum_vp(v)^2e(v)
 &=b_x\langle p,e\rangle+\sum_vp(v)(p(v)-b_x)e(v)\\
 &\le b_xD/2+\sqrt{M G_pD}.
\end{align*}
Here Cauchy--Schwarz uses
$\sum_vp(v)^2(p(v)-b_x)^2\le M\sum_vp(v)(p(v)-b_x)^2=M G_p$.
Moreover, $\sum_vp(v)e(v)^2\le MD$ and $\sum_ve(v)^3\le\|e\|_2^3\le D^{3/2}$. Expanding $(p+e)^3$ and using $a^2\ge b_x^2$ yields
\[
 F_3(r)-a^2
 \le G_p+\tfrac32b_xD+3\sqrt{MG_pD}+3MD+D^{3/2}.
\]
Since $M,b_x=O(b)$, $D\le\delta b$, and $G_p=O(\delta b^2)$, the right-hand side is $O(\delta b^2+\delta^{3/2}b^{3/2})$, proving the stated bound.
\end{proof}

\begin{lemma}[Comparison at one start]\label{lem:restricted-comparison}
Fix $s$ in a component $C$ and $0<\theta\le1/64$. Use two independent batches of $q=\lceil K\sqrt n/\theta\rceil$ endpoints, at lengths $t$ and $2t$ from $s$, to form $U=\widehat F_2(P^t(s,\cdot))$ and $V=\widehat F_2(P^{2t}(s,\cdot))$, where $K$ is a sufficiently large absolute constant. Accept if $U\le(1+4\theta)V$. If $\Delta_C\le\theta$, acceptance has probability at least $9/10$. If $C$ is walk-regular and $\Delta_C\ge8\theta$, rejection has probability at least $9/10$.
\end{lemma}

\begin{proof}
Write $b=b_C$, $\delta=\Delta_C$, and $Z=U-(1+4\theta)V$. For $\delta\le1$, independence, Eq.~\eqref{eq:refined-variance}, and Lemma~\ref{lem:walk-third-moment} give
\[
 \Var Z\le C\left(\frac b{q^2}+\frac{\delta b^2}q
                         +\frac{\delta^{3/2}b^{3/2}}q\right).
\]
If $\delta\le\theta$, then $\E Z\le-3\theta b$. Dividing the variance by $\theta^2b^2$, the three terms are at most constant multiples of $K^{-2}$, $(K\sqrt n)^{-1}$, and $\sqrt\theta/(K\sqrt{nb})$, respectively. Since $b\ge1/n$, Chebyshev's inequality proves acceptance with probability at least $9/10$ for large enough $K$.

For rejection, walk-regularity makes $A_s=(1+\delta)b$ and $B_s=b$, so $\E Z=(\delta-4\theta)b\ge\delta b/2$. If $8\theta\le\delta\le1$, the variance divided by $\delta^2b^2$ is at most
\[
 C\left(\frac1{q^2b\delta^2}+\frac1{q\delta}
                         +\frac1{q\sqrt b\sqrt\delta}\right)=O(1/K).
\]
If $\delta>1$, use $F_3(P^t(s,\cdot))\le((1+\delta)b)^{3/2}$ and
$F_3(P^{2t}(s,\cdot))-b^2\le\delta b^2$, the latter following from Eq.~\eqref{eq:appendix-pointwise}. The normalized variance is then at most
\[
 C\left(\frac{1+\delta}{q^2b\delta^2}
       +\frac{(1+\delta)^{3/2}}{q\sqrt b\delta^2}
       +\frac1{q\delta}\right)=O(1/K).
\]
Chebyshev's inequality proves the rejection guarantee in both cases.
\end{proof}

\begin{proof}[Proof of Proposition~\ref{prop:restricted-tester}]
Use $\eta,t$ and the constant $a_0$ from the proof of Theorem~\ref{thm:main}. At each sampled start, apply Lemma~\ref{lem:restricted-comparison} with $\theta=\eta/4$, amplify by independent repetition, and reject if any amplified comparison rejects. Use the same $O_{d,c,\beta}(1/\eps)$ uniform starts as in the unrestricted tester.

For completeness, Lemma~\ref{lem:mixing} gives $\Delta_C\le\eta/10<\theta$ in every expanding component, whether or not it is walk-regular. For soundness on a componentwise walk-regular graph, $A_v$ is constant on each component, so $D_s$ is empty. A start is therefore good exactly when $\Delta_{C(s)}\le2\eta$. The repair argument in the proof of Theorem~\ref{thm:main} shows that an $\eps$-far graph has an $\Omega_{d,c,\beta}(\eps)$ fraction of starts with $\Delta_{C(s)}>2\eta=8\theta$. The amplified comparison detects any such start with high probability.

Each comparison costs $O(\sqrt n\,t/\eta)$ queries before amplification. The identity
\[
 1/\eta=a_0^{-1}\max\{\eps^{-1},n^{\beta/3}\}
\]
gives the stated query bound. If the whole graph is walk-regular, $A_s/B_s$ is independent of $s$. If this common ratio were at most $1+2\eta$, every start would be good and the same repair bound would contradict farness. Thus every start of a far walk-regular graph fails the comparison condition. One start with constant amplification suffices.
\end{proof}

\end{document}